\RequirePackage{amsthm}
\documentclass[pdflatex,sn-basic,Numbered,]{sn-jnl}

\usepackage{graphicx}%
\usepackage{multirow}%
\usepackage{amsmath,amssymb,amsfonts}%
\usepackage{mathrsfs}%
\usepackage[title]{appendix}%
\usepackage{xcolor}%
\usepackage{textcomp}%
\usepackage{manyfoot}%
\usepackage{booktabs}%
\usepackage{algorithm}%
\usepackage{algorithmicx}%
\usepackage{algpseudocode}%
\usepackage{listings}%
\usepackage{anyfontsize}

\theoremstyle{thmstyleone}%
\newtheorem{theorem}{Theorem}
\newtheorem{proposition}[theorem]{Proposition}%
\newtheorem{definition}[theorem]{Definition}%
\newtheorem{example}[theorem]{Example}%

\theoremstyle{thmstyletwo}%

\theoremstyle{thmstylethree}%

\usepackage{xcolor}
\usepackage{enumerate}
\usepackage{enumitem}
\usepackage{proof}
\usepackage[all]{xy}
\newcommand{\seqcomp}{\mathbin{;}}

\renewcommand{\phi}{\varphi}
\renewcommand{\epsilon}{\varepsilon}

\newcommand{\declaration}[1]{\noindent \textbf{Declaration:} #1 \bigskip}
\newcommand{\acknowledgements}[1]{\noindent \textbf{Acknowledgements:} #1 \bigskip}
\newcommand{\conflict}[1]{\noindent \textbf{Competing Interests:} #1 \bigskip}

\renewcommand{\emptyset}{\varnothing}

\newcommand{\proves}{\vdash}

\newcommand{\partialto}{\rightharpoonup}

\newcommand{\seq}{\triangleright}

\newcommand{\calculus}[1]{\ensuremath{\mathsf{#1}}}

\newcommand*{\setStyle}[1]{\ensuremath{\mathbb{#1}}}
\newcommand{\setGoals}{\setStyle{GOALS}}
\newcommand{\setRed}{\setStyle{RED}}
\newcommand{\setEvents}{\setStyle{EVENTS}}
\newcommand{\logic}[1]{\mathscr{#1}}

\newcommand{\func}[1]{\mathcal{#1}}
\newcommand{\funcList}{\textsc{list}}

\newcommand{\funcfinPower}{\mathcal{P}_{<\omega}}

\newcommand{\llbracket}{[\![}
\newcommand{\rrbracket}{\ensuremath{]\!]}}

\begin{document}

\title[Semantic Foundations of Reduction Reasoning]{Semantic Foundations of Reductive Reasoning}


\author*[1,2]{\fnm{Alexander V.} \sur{Gheorghiu}}\email{a.v.gheorghiu@soton.ac.uk}

\author[2,3,4]{\fnm{David J.} \sur{Pym}}\email{d.pym@ucl.ac.uk}

\affil*[2]{\orgdiv{School of Electronics and Computer Science}, \orgname{University of Southampton}, \orgaddress{\street{University Street}, \city{Southampton}, \postcode{SO17 1BJ}, 
\country{United Kingdom}}}

\affil*[1]{\orgdiv{Department of Computer Science}, \orgname{University College London}, \orgaddress{\street{Gower Street}, \city{London}, \postcode{WC1E 6EA}, 
\country{United Kingdom}}}

\affil*[3]{\orgdiv{Department of Philosophy}, \orgname{University College London}, \orgaddress{\street{Gordon Square}, \city{London}, \postcode{WC1H 0AW}, 
\country{United Kingdom}}}

\affil*[4]{\orgdiv{Institute of Philosophy}, \orgname{University of London}, \orgaddress{\street{Malet Street}, \city{London}, \postcode{WC1E 7HU}, 
\country{United Kingdom}}}


\abstract{
 The development of logic has largely been through the \emph{deductive} paradigm: conclusions are inferred from established premisses. However, the use of logic in the context of both human and machine reasoning is typically through the dual \emph{reductive} perspective: collections of sufficient premisses are generated from putative conclusions. We call this paradigm, \emph{reductive logic}. This expression of logic encompass as diverse reasoning activities as proving a formula in a formal system to seeking to meet a friend before noon on Saturday. This paper is a semantical analysis of reductive logic. In particular, we  provide mathematical foundations for representing and reasoning about \emph{reduction operators}. Heuristically, reduction operators may be thought of as `backwards' inference rules. In this paper, we address their mathematical representation, how they are used in the context of reductive reasoning, and, crucially, what makes them \emph{valid}. 
}




\maketitle
\vspace{-10mm}

\declaration{
This work has been partially supported by the UK EPSRC grants EP/S013008/1 and EP/R006865/1 and by the EU MOSAIC MCSA-RISE project.}

\acknowledgements{
We are grateful to Simon Docherty, Tao Gu, and Eike Ritter for discussions of various aspects of this work.
}

\conflict{
The authors have no competing interests to declare that are relevant to the content of this article.
}

\section{Introduction} \label{sec:introduction}

The definition of a system of logic may be given \emph{proof-theoretically} as a collection of rules of inference that, when composed in specified ways, determine \emph{proofs}; that is, formal constructions that establish that a conclusion is a consequence of some assumptions or axioms. In other words, proofs are objects regulated by rules from a given formal system that determine that an inference of a conclusion from a collection of premisses has been established:
\[
\frac{\mathrm{Established \; Premiss}_1 \quad \ldots \quad \mathrm{Established \; Premiss}_k}{\mathrm{Conclusion}}{\big\Downarrow} \vspace{2mm}
\]
We call this proof-theoretic formulation \emph{deductive logic}.

Deductive logic is useful as a way of defining what proofs are, but it does not reflect either how logic is typically used in practical reasoning problems or the method by which proofs are found. In practice, proofs are typically constructed by starting with a desired, or putative, conclusion and applying the rules of inference \emph{backwards}. Read from conclusion to premisses, the rules are sometimes called \emph{reduction operators}, and denoted
\[
\frac{\mathrm{Sufficient \; Premiss}_1 \quad \ldots \quad \mathrm{Sufficient \; Premiss}_k}{\mathrm{Putative \; Conclusion}} {\big\Uparrow}  \vspace{2mm}
\]
We call the constructions in a system of reduction operators \emph{reductions}.  This proof-theoretic formulation has been dubbed \emph{reductive logic}~\cite{pym2004reductive}. 

Note that reductive logic is not \emph{a} logic in the sense of, for example, classical, intuitionistic, or modal logic, but a conception of logic in the sense of \emph{deductive}, \emph{inductive}, or \emph{abductive} logic. 


To clarify, reductive logic is defined with respect to a formal specification of a logic and does not encompass semantic reasoning in general. Specifically, the figures above illustrate particular inferences within formal systems, rather than arbitrary steps of problem-solving. As such, reductive logic represents a relatively narrow domain within the broader field of analysis~\cite{sep-analysis}. More precisely, it pertains to the aspects of analysis employed in computer-assisted formal reasoning, as we discuss below.

Also to avoid confusion, note that reductive logic should not be conflated 
with the superficially related concept of reverse mathematics, as developed by Friedman~\cite{Friedman1975} and others. Reverse mathematics investigates which axioms, within a given formalization of mathematics, are necessary for specific theorems to hold. By contrast, reductive logic focuses on the proof theory and semantics underlying the process of finding proofs --- via reduction and 
search --- within fixed systems of logic. 


Focusing on areas with formal, well-defined notions of inference allows us to develop a more concrete semantic theory of reductive reasoning. Specifically, our aim is to create a semantics of reductions that not only identifies complete proofs but also clarifies the meaning of unfinished or incomplete proof-searches. This approach places our investigation squarely within the domain of proof-theoretic semantics (P-tS), which seeks to explain the validity of proofs in formal logic. So far, P-tS has largely been developed in the context of deductive logic. We have argued that it is a natural and metaphysically well-motivated semantic framework that is explanatory for both reductive and deductive logic. We therefore argue that this suggests that these two views of formal logic are as valid as each other as ways of modelling reasoning. 

For a comprehensive treatment of P-tS, we refer to Schroeder-Heister~\cite{Schroeder2006validity,Schroeder2007modelvsproof}, though we provide the key highlights here. While P-tS is typically framed in the context of deductive logic, its core principles are directly applicable to reductive reasoning. In the Dummett-Prawitz tradition of P-tS --- perhaps the most extensively studied --- direct proofs (closed, normalized proofs) are considered canonically valid, whereas indirect proofs rely on transformations to achieve directness. This distinction introduces a technical separation between proof structures, which represent potential proofs, and justifications, which define the transformations that validate them. Similarly, in reductive reasoning, several steps of reduction may yield a potential proof or contribute proof-theoretic content, even if additional manipulation and transformation are required to achieve the final proof of the intended goal --- that is, the putative conclusion one desires to prove.

Na\"{\i}vely, reductive logic may appear simply as deductive logic read backwards. This heuristic suffices in many cases and is useful for understanding deductive logic; for example, proof of cut-elimination for sequent calculi (see Gentzen~\cite{Gentzen}) can be understood from this perspective. However, the space of reductions studied in reductive logic is larger than the space of proofs, as studied in deductive logic: while the reductive system can simply reconstruct all the correct proofs in the underlying system, it also allows choices to be made that yield reductions that do not correspond to proofs. For example, when no further reduction operators can be applied to a premiss and that premiss is not an axiom of the logic. Reductive logic is thus a paradigm of logic in parallel to deductive logic and cannot be defined purely in terms of it.

While the restriction to proof systems may seem limiting, the conception of reductive logic in this paper is sufficiently general to encompass reasoning activities as diverse as proving a formula in a formal system and seeking to meet a friend before noon on Saturday. Importantly, it more closely resembles what mathematicians do when proving theorems and, more generally, how people solve problems using formal representations. It is also the paradigm of logic used for diverse applications in informatics and other systems-oriented sciences, including, but not limited to, areas such as program and system verification, natural language processing, and knowledge representation. It is the reductive perspective that underpins the use of logic in proof-assistants --- for example, LCF \cite{Gordon1979}, HOL \cite{Gordon1993}, Isabelle \cite{Paulson1994}, Coq \cite{Inria,Bertot2004}, Twelf \cite{Twelf,Pfenning1999}, and more~\cite{wikiproofassistants}. More generally, it is a restricted model of human reasoning, which is discussed extensively, for example, in the work of Kowalski~\cite{Kowalski1986} and Bundy~\cite{Bundy1983}. We discuss this further in Section~\ref{sec:reductive-logic-1}. 
 
We seek a semantic foundation for reduction operators. Our motivation is both philosophical and pragmatic. Philosophically, we recognize that reductive logic is a paradigm through which much practical problem solving occurs, as discussed above (examples below). Pragmatically, the increasing significance of technologies reliant on reductive logic suggests the pressing need for a more comprehensive and mathematically more developed semantic theory than currently available --- see, for example, Pym and Ritter~\cite{pym2004reductive}. Hence, in particular, we seek a representation of reduction operators as functions encapsulating the action suggested by the picture above,
\[
\text{Putative Conclusion} \quad \longrightarrow \quad \text{Sufficient Premiss}_1\,, \ldots , \,\text{Sufficient Premiss}_n
\]
Having done this, we can then analyze them and formally study reductive reasoning. It is not only the representation of reduction operators as mathematical objects that is important: we seek an account of what makes a reduction operator and its deployment \emph{valid}.

Historically, semantics in logic has been dominated by the paradigm of \emph{denotationalism}, where the meaning of logical structures is given in terms of abstract interpretations. For instance, in model-theoretic semantics, propositions are interpreted relative to abstract algebraic structures called \emph{models} (see Schroeder-Heister~\cite{Schroeder2007modelvsproof}). This paradigm is suitable for \emph{deductive} logic. However, other paradigms of logic may benefit from different approaches.

For \emph{reductive} logic, denotationalist accounts have been useful and thoroughly explored (see Pym and Ritter~\cite{pym2004reductive},  Komendantskaya et al.~\cite{Komendantskaya2010, Komendantskaya2011, Komendantskaya16}, and Gheorghiu et al.~\cite{GDP2023bi-lp}). These studies are discussed below. In this paper, we generalize these results by removing auxillary and domain-specific aspects to concentrate on the semantics of reductive reasoning itself.

We also examine semantics with a \emph{proof-theoretic} focus, this is the connection to P-tS. That is, we give validity conditions not in terms of abstract algebraic structures, but in terms of what `proofs' are being witnessed during reductive reasoning. Pym and Wallen~\cite{PW1993logic} provided such an account in terms of \emph{proof-valued computations}.  

To this end, we draw substantively upon Milner's theory of \emph{tactical proof}~\cite{milner1984tactics} in Section~\ref{sec:reductive-logic-2}. The historical and technical significance of tactical proof in the context of computer-assisted formal reasoning is profound; it forms the basis for almost all proof assistants (see Gordon~\cite{gordon2015tactics}). Ultimately, this inspires the P-tS account of reductive logic in this paper.

We begin, in Section~\ref{sec:semantics-of-proofs}, by explicating the notion of reduction and its current semantics in comparison to well-established techniques in deductive logic. This helps expose the subtlety and challenges involved in reductive logic. In Section~\ref{sec:reductive-logic-1}, we provide a denotational semantics of the reduction operators and soundness and completeness conditions that determine the validity of a set of reduction operators. In Section~\ref{sec:reductive-logic-2}, we give an operational account of reduction operators in terms of structural operational semantics that enables a more refined account of their use and validity. In Section~\ref{sec:control}, we discuss future problems for a satisfactory foundation of reductive reasoning; viz., the expression and semantics of strategies for proof-search. This moves us from a semantic foundation of reduction operators, which we may think of as the \emph{local} perspective on reduction reasoning, to a full semantic foundation of reductive logic, the corresponding \emph{global} perspective on reductive logic. The paper ends in Section~\ref{sec:conclusion} with a summary of the results obtained.

This paper should be seen as a companion to \cite{GPTactical}, in which P-tS is used to provide a semantics for tactical proof, mentioned above. By contrast, this paper provides a general foundation for reductive reasoning in terms of P-tS  and obtains tactical proof as a significant instance. Happily, the two analyses are mutually coherent. \\

\medskip \noindent \textbf{Technical Background.} We assume familiarity with formal logic (see, for example, van Dalen~\cite{vanDalen}) and basic proof theory (see, for example, Troelstra and Schwichtenberg~\cite{troelstra2000basic} and Negri~\cite{Negri2005}). We use $\phi, \psi, \chi, \ldots$ to denote logical formulae and $\Gamma, \Delta, \ldots$ to denote collections thereof. We use $\seq$ as the sequent symbol and we distinguish it from the consequence judgement $\vdash$ of a logic; that is, a sequent is an expression of an assertion in a logic, which may be valid or invalid, but a consequence is an assertion that is valid.

For the technical portions of this paper, we will assume some basic concepts and terminology from category theory --- see, for example, Cheng~\cite{cheng2022joy} --- in particular, categories, functors, adjunctions, and natural transformations. However, these technical details are not essential for the thesis, they simply give a systematic mathematical treatment of the ideas presented. As we are always working in the category of sets and (partial) functions, relatively little background is required beyond typical mathematical training. 

We use $\funcfinPower$ to denote the \emph{finite} powerset functor --- that is, given a set $X$, we write $\funcfinPower(X)$ to denote the set of all finite subsets of $X$ --- and $\funcList$ to denote the list functor --- that is, given a set $X$, we write $\funcList(X)$ to denote the set of all finite lists of $X$. We will use the concept of \emph{$F$-algebra} and \emph{$F$-coalgebra} (for a functor $F$) in various parts of this paper, but they are just functions.\footnote{$F$-algebras generalize the notion of algebraic structure. If $\mathcal{C}$ is a category and $F : \mathcal{C} \rightarrow \mathcal{C}$ is an endofunctor, then an $F$-algebra is a pair $(A,\alpha)$, where $A$ is an object of $\mathcal{C}$ and $\alpha : F(A) \rightarrow A$ is a morphism in $\mathcal{C}$. $A$ is called the carrier of the algebra. If we take $\alpha : A \rightarrow F(A)$, then we have an $F$-coalgebra. For endofunctors $F$ on the category of sets, $F$-algebras are functions on the carrier set and $F$-coalgebras are countably infinite streams (sequences) over the carrier set. Often, we refer to the maps $\alpha$ as (co)algebras.} The nomenclature of category theory enables us to apply well-established techniques and constructions. For a general introduction to coalgebra, see, for example, Rutten~\cite{RUTTEN20003} and Jacobs~\cite{jacobs2017introduction}.

We use basic co-induction in some definitions and proofs and, accordingly, give a brief explanation of how it works presently; for a more detailed discussion, see, for example, Barwise and Moss~\cite{Barwise1996} and Rutten~\cite{RUTTEN20003}. We have to use co-induction (as opposed to induction) because reduction spaces are co-recursively generated and so may be infinite objects. We follow Reitzig~\cite{Reitzig} for a succinct explanation of co-induction as the dual of induction.

In an \emph{inductive} definition of a set, one begins with some \emph{base} elements and applies \emph{constructors} that say how one builds new elements of the set from existing ones. For example, a set of finite sequences, or `strings', $\mathbb{S}$ may be defined by the following:
\[
\begin{cases}
    \epsilon \in \mathbb{S} & \text{Base Case} \\
    w \in  \mathbb{S} \Longrightarrow aw \in  \mathbb{S}  & \text{Condition 1} \\
    aw \in  \mathbb{S} \Longrightarrow baw \in  \mathbb{S} & \text{Condition 2}
\end{cases}
\]
The set $\mathbb{S}$ inductively defined in this way is the \emph{smallest} set satisfying these conditions. Hence, as defined above, $\mathbb{S}$ is the set of strings of $a$ and $b$ with no two subsequent $b$s. This is a perfectly adequate account of induction, but one can give a more subtle treatment that explains precisely how it works in terms of fixed points.

The definition above determines the following map $f$ on sets of strings,
\[
f(X) := X \cup \{\epsilon\} \cup \{aw \mid w \in X\} \cup  \{baw \mid aw \in X\}
\]
Observe that $f$ is monotone and the sets of strings form a complete lattice under set inclusion. Hence, the Knaster-Tarski Theorem~\cite{Knaster1928,Tarski1955} tells us that the set of fixed points for $f$ forms a complete lattice. The defined set $\mathbb{S}$ is the \emph{least} fixed point; that is, the smallest set $X$ such that $f(X)=X$. Since $f$ is continuous, $\mathbb{S}$ is the colimit of the following $\omega$-chain,
\[
\xymatrix{
\emptyset \ar[r] &  \underbrace{\{\epsilon\} \cup \{aw\}}_{\emptyset \cup f(\emptyset)}  \ar[r] &  \underbrace{\{\epsilon, a\} \cup f(\{\epsilon, a\})}_{\emptyset \cup f(\emptyset) \cup f(f(\emptyset))} \ar[r] & \ldots \ar[r] & \underbrace{\mathbb{S}}_{\bigcup_n^\omega f^n(\emptyset)}
}
\]
Here each arrow denotes an application of $f$ and $\mathbb{S}$ denotes the colimit of the chain.

For a \emph{co-inductive} definition, we use \emph{destructors} rather than constructors. Intuitively, we reverse the implications in the inductive definition; observe that in doing so the `base case' becomes vacuous. That is, rather than say `if $w$ is included, then so is $aw$', we say `if $aw$ is included, then so was $w$'. This determines a set $\mathbb{S}'$ with the following property: we can take arbitrarily long prefixes away from any word in $\mathbb{S}'$ and remain in $\mathbb{S}'$. Observe that this property does not hold for finite strings, hence the co-inductive definition requires us to consider \emph{infinite} strings.

While induction determines the least fixed point satisfying $f$, the co-inductive definition determines the (unique) \emph{greatest} fixed point --- that is, the \emph{largest} set $X$ such that $f(X)=X$. Thus $\mathbb{S}'$ contains all \emph{infinite} strings of $a$ and $b$ with no consecutive occurrences of $b$s. In other words, it is the limit of any $\omega^{\text{op}}$-chain,
\[
\xymatrix{
1 & \ar[l] f(1) & \ar[l] f(f(1)) & \ar[l] f(f(f(1))) & \ar[l] \ldots
}
\]
in which $1$ is a singleton (i.e., a terminal object in the category), and the arrows denote the duals of the arrows in the least fixed point case (i.e., the unique morphisms that remove a prefix from a string). This is quite intuitive: we cannot construct $\mathbb{S}'$ by an iterative process on elements as its elements are infinite, so instead we determine it by starting with all strings and removing those that do not satisfy the conditions after an arbitrary amount of time.

This concludes the background on co-induction.

\section{The Semantics of Reductive Logic} \label{sec:semantics-of-proofs}

In this section, we motivate the idea of reductive logic and explicate its current semantics. We set this discussion in the historical development of \emph{intuitionistic logic} (IL)~\cite{SEP-IPL} as it is here that the ideas are well known and are the most developed.  Subsequently, in  Section~\ref{sec:reductive-logic-1} and Section~\ref{sec:reductive-logic-2}, we address the semantics of reduction operators, the things that generate reductions.

In logic, we have structures called \emph{arguments} that represent evidence for a consequence of a logic --- for example, natural deduction arguments after Gentzen~\cite{Gentzen}, after 
 Fitch~\cite{fitch1952symbolic}, after Lemmon~\cite{lemmon1978beginning},  and so on. When these arguments are \emph{valid} they represent a \emph{proof} of some statement. The semantics of proofs for particular logics have been very substantially developed for particular logics. Perhaps the best examples is the BHK interpretation of \emph{intuitionism}.
 
 Intuitionism, as defined by Brouwer~\cite{brouwer1913intuitionism}, is the view that an argument is valid when it provides sufficient evidence for its conclusion. This is IL. Famously, as a consequence, IL rejects \emph{the law of the excluded middle} --- that is, the meta-theoretic statement that either a statement or its negation is valid. This law is equivalent to the principle that, in order to prove a proposition, it suffices to show that its negation is contradictory. In IL, such an argument does not constitute sufficient evidence for its conclusion. 
 
Heyting~\cite{heyting1966intuitionism} and Kolmogorov~\cite{kolmogorov} provided a semantics for intuitionistic proof that captures the evidential character of intuitionism, called the Brouwer-Heyting-Kolmogorov (BHK) interpretation of IL. It is now the standard explanation of the logic --- see, for example, van Atten~\cite{SEP-IPL}. 

The \emph{propositions-as-types} correspondence --- see Howard, Barendregt, and others \cite{howard1980formulae,BDS2013,Barendregt1991,Barendregt1993} --- gives a standard way of instantiating the denotations of proofs in the BHK interpretation of \emph{intuitionistic propositional logic} (IPL) as terms in the simply-typed 
$\lambda$-calculus. Technically, the set-up can be sketched as follows: a judgement 
that $\mathcal{D}$ is an $\calculus{NJ}$-proof of the sequent $\phi_1 , \ldots , \phi_k \seq \phi$ corresponds to a typing 
judgement 
\[
    x_1 : A_1 , \ldots , x_k : A_k \vdash M(x_1,\ldots, x_k) : A 
\]
where the $A_i$s are types corresponding to the $\phi_i$s, the $x_i$s correspond to placeholders for proofs of the $\phi_i$s, the $\lambda$-term $M(x_1 , \ldots, x_k)$ corresponds to $\mathcal{D}$, and 
the type $A$ corresponds to $\phi$. 

Lambek~\cite{lambek1980lambda} gave a more abstract account by showing that the simply-typed $\lambda$-calculus is the internal language of \emph{cartesian closed categories} (CCCs), thereby giving a categorical semantics of proofs for IPL. In this set-up, 
a morphism 
\[
    \llbracket \phi_1 \rrbracket \times \ldots \times \llbracket \phi_k \rrbracket 
        \stackrel{\llbracket \mathcal{D} \rrbracket}{\longrightarrow} 
            \llbracket \phi \rrbracket
\]
in a CCC, where $\times$ denotes cartesian product, that interprets the $\calculus{NJ}$-proof $\mathcal{D}$ of $\phi_1 , \ldots , \phi_k \seq \phi$ also interprets the 
term $M$, where the $\llbracket \phi_i \rrbracket$s interpret also the $A_i$s and $\llbracket \phi \rrbracket$ also interprets 
$A$.

\begin{figure}[ht]
    \hrule
    \[
\xymatrix{
    x:\Gamma \proves M(x):\phi \ar@{<->}[rr] & &   \llbracket \Gamma \rrbracket \stackrel{\llbracket \mathcal{D} \rrbracket}{\rightsquigarrow} \llbracket \phi \rrbracket \\
    & \ar@{<->}[ul] \mathcal{D} \Rightarrow \Gamma \seq \phi \ar@{<->}[ur] &
}
\]
\hrule \vspace{1mm}
    \caption{Curry-Howard-Lambek Correspondence}
    \label{fig:chl}
\end{figure}

Altogether, this describes the \emph{Curry-Howard-Lambek} correspondence for IPL. It may be summarized by Figure~\ref{fig:chl} in which:
\begin{itemize}[label=--]
    \item $\mathcal{D} \Rightarrow \Gamma \seq \phi$ denotes that $\mathcal{D}$ is an  $\calculus{NJ}$-derivation of $\phi$ from $\Gamma$;
    \item $x:A_\Gamma \vdash M(x):A_\phi$ denotes a typing judgment, as described above, corresponding to $\mathcal{D}$; and,
    \item $\llbracket \Gamma \rrbracket \stackrel{\llbracket \mathcal{D} \rrbracket}{\rightsquigarrow} \llbracket \phi \rrbracket$ denotes that $\llbracket \mathcal{D} \rrbracket$ is a morphism from  $\llbracket \Gamma \rrbracket$ to $\llbracket \phi \rrbracket$ in a CCC. 
\end{itemize} 

\vspace{2mm}
To generalize to full IL (and beyond), Seely~\cite{seely1983hyperdoctrines} modified this categorical set-up and introduced \emph{hyperdoctrines} --- indexed categories of CCCs with coproducts over a base with finite products. Martin-L\"of~\cite{martin1975intuitionistic} gave a formulae-as-types correspondence for predicate logic using dependent type theory. Barendregt~\cite{Barendregt1991} gave a systematic treatment of type systems and the propositions-as-types correspondence. A categorical treatment of dependent types came with Cartmell~\cite{cartmell} --- see also, for examples among many, work by Streicher~\cite{Streicher1988}, Pavlovi\'c~\cite{Pavlovic1990}, Jacobs~\cite{Jacobs}, and Hofmann~\cite{Hofmann1997}. In total, this gives a semantic account of \emph{proof} for first- and higher-order predicate intuitionistic logic based on the BHK interpretation. 

This is all very well for explaining what is a proof in IL. In reductive logic, however, the space of objects considered contains also things that are not proofs and cannot be continued to form proofs. Pym and Ritter~\cite{pym2004reductive} have provided a general semantics of reductive logic in the context of classical and intuitionistic logic through polynomial categories; that is, by extending the categories in which arrows denote proofs for a logic by additional arrows that supply `proofs' for propositions that do not have proofs but appear during reduction. This generalization of the BHK interpretation is known as the \emph{constructions-as-realizers-as-arrows correspondence} in Figure~\ref{fig:cor}:
\begin{enumerate}[label=--]
    \item \mbox{$\Phi \Rightarrow \Gamma \seq \phi$} denotes that $\Phi$ is a sequence of reductions for the sequent $\Gamma \seq \phi$;
    \item $[\Gamma] \proves [\Phi] : [\phi]$ denotes that $[\Phi]$ is a \emph{realizer} of [$\phi$] with respect to the assumptions $[\Gamma]$; and,
    \item  $\llbracket \Gamma \rrbracket \stackrel{\llbracket \Phi \rrbracket}{\rightsquigarrow} \llbracket \phi \rrbracket$ denotes that $\llbracket \Gamma \rrbracket$ is a morphism from  $\llbracket \Gamma \rrbracket$ to $\llbracket \phi \rrbracket$ in the appropriate polynomial category. 
\end{enumerate}

They also defined a judgement 
$w \Vdash_\Theta (\Phi:\phi)\Gamma$  which says that $w$ is a world witnessing that $\Phi$ is a reduction of $\phi$ to $\Gamma$, relative to the indeterminates of $\Theta$.

What has just been presented gives a semantics of \emph{reductions} that explicates how they relate to proofs and certify the validity of sequents. In this paper, we turn to the semantics of the reduction operators themselves; that is, the things which generate the reductions. Unlike rules, which specify particular transformations, reduction operators contain a certain degree of non-determinism rendering the subject more subtle.

\begin{figure}[ht]
    \hrule
    \[
\xymatrix{
    [\Gamma] \proves [\Phi]:[\phi] \ar@{<->}[rr] & &   \llbracket \Gamma \rrbracket \stackrel{\llbracket \Phi \rrbracket}{\rightsquigarrow} \llbracket \phi \rrbracket \\
    & \ar@{<->}[ul] \Phi \Rightarrow \Gamma \seq \phi \ar@{<->}[ur] &
}
\]
\hrule \vspace{1mm}
\caption{Constructions-as-Realizers-as-Arrows Correspondence}
    \label{fig:cor}
\end{figure}

\section{Reduction Operators I --- Denotational Semantics} \label{sec:reductive-logic-1}

In this section, we give a \emph{denotational} semantics of reduction operators beginning from first principles. We then use this semantics to express the space of reductions.

\subsection{Denotational Semantics}
As described in Section~\ref{sec:introduction}, reductive logic is concerned with `backwards' inference,
\[
\frac{\mathrm{Sufficient \; Premiss}_1 \quad \ldots \quad \mathrm{Sufficient \; Premiss}_k}{\mathrm{Putative \; Conclusion}} {\big\Uparrow}  \tag{$\dagger$}
\]
This framework not only describes reduction in a formal system but also applies to reductive reasoning in more general settings.

\begin{example}[see Milner \cite{milner1984tactics}] \label{ex:Alice-Bob-1}
Two friends, Alice and Bob, desire to meet before noon on Saturday. We shall model how they may reason about accomplishing this goal.

Let $G$ be the statement that Alice and Bob did meet before noon on Saturday. For $G$ to hold, Alice and Bob reason that they need to arrive at Waterloo Station before noon on Saturday. That is, it suffices for $G_1$ and $G_2$ to hold:
\begin{itemize}[label=--]
    \item $G_1$: Alice arrives under the clock at Waterloo Station before noon on Saturday
    \item $G_2$: Bob arrives under the clock at Waterloo Station before noon on Saturday
\end{itemize}
This represents a reduction of the putative conclusion $G$ to the sufficient premisses $G_1$ and $G_2$.
\end{example}

The precise meaning of $(\dagger)$ has been formalized by Pym and Ritter~\cite{pym2004reductive}, though the earliest account is perhaps due to Kleene~\cite{kleene2013mathematical}. Fixing a logic $\logic{L}$, the ideas are as follows:

\begin{enumerate}[label=R\arabic*,font=\bfseries]
    \item We assume that our logic $\logic{L}$ comes with a proof system. Each inference rule denotes a \emph{reduction operator} mapping assertions in $\logic{L}$ to lists of assertions in $\logic{L}$. \label{def:r1}
    \item The putative conclusion is an assertion in our logic $\logic{L}$, called the \emph{goal}. \label{def:r2}
    \item The assertions that must be proved in order to have a proof system of the initial assertion, called \emph{subgoals}, are given by the corresponding instance of the sufficient premisses of a reduction operator. \label{def:r3}
\end{enumerate}

While \ref{def:r2} and \ref{def:r3} merely specify the relationship between the terms involved, \ref{def:r1} does more than needed. In practice, it is clearest to begin with a proof system and read the inference rules backwards to generate an adequate set of reduction operators, but this is inessential and limiting for defining reductive logic for a given logic. Pym and Ritter~\cite{pym2004reductive} acknowledge that much of their analysis can be recovered purely semantically. Indeed, Gheorghiu and Pym~\cite{GP2023acs,GP2023semantical} have shown that reductive logic bridges semantics and proof theory. For example, in \cite{GP2023semantical}, it is shown for the bunched logic BI (and so, implicitly, for intuitionistic propositional logic also) that a bisimulation between the transition system generated by reductive proof and the one generated by the satisfaction relation of its relational model-theoretic semantics can be used to establish completeness.


The point of \ref{def:r1} is twofold: it ensures the \emph{validity} of the reduction operators, and it ensures their correctness --- that is, if the putative conclusion is indeed valid, then it will eventually be reduced to trivial premisses. Thus, it says more than what ($\dagger$) intends. Presently, let us drop \ref{def:r1} in place of the following:
\vspace{2mm}
\begin{enumerate}[label=R\arabic*$'$,font=\bfseries]
    \item We have a finite set of reduction operators $\setRed$ for assertions in the language of $L$, which map assertions to sets of assertions \label{def:r1prime}
\end{enumerate}
\vspace{2mm}
The assumption that the set is finite enables the mathematical treatment and is quite representative as proof systems typically have finitely many rule schemas.

Having made this simplification in the overhead, we no longer need to fix a logic $\logic{L}$ but only an assertion language over which reduction is taking place. The validity of reductions in this setup may then be determined by their correctness relative to a logical system, as in the standard practice of semantics.

With heuristics \ref{def:r1prime}, \ref{def:r2}, and \ref{def:r3} in place, we can give a precise, mathematical account of $(\dagger)$.

Let $\setGoals$ denote the set of all assertions in the language $\mathcal{L}$. The heuristics above suggest that a reduction operator is a partial function from goals to finite sets of goals:
\[
\rho: \setGoals \rightharpoonup \funcfinPower(\setGoals)
\]

\begin{example}[Example~\ref{ex:Alice-Bob-1} cont'd]
Alice's and Bob's reasoning is modelled by the following reduction,
\[
\rho: G \mapsto \{G_1, G_2\}
\]
\end{example}

Such reduction operators describe one step of reduction. Reasoning reductively means chaining together such steps, reducing the subgoals to further subgoals. This introduces the idea of \emph{state} as the set of goals to be reduced.

Initially, the state is a singleton consisting of a single goal, the putative conclusion. Subsequently, it consists of a finite set of (sub)goals generated so far. Mathematically, a reduction operator $\rho$ moves a state $S$ to a next state $S'$ by replacing one of the goals with a list of subgoals,
\[
S  = \{G_1, \ldots ,G_n\} \overset{\rho}{\longrightarrow}  (\{G_1, \ldots ,G_n\}\setminus\{G_i\})\cup \rho(G_i) = S'
\]
Observe that a sequence of reductions may go on infinitely or terminate. When it terminates, one of the following two things can happen: 

\begin{enumerate}[label = \textbf{T\arabic*}]
\item There are no further subgoals --- that is, one has reduced all the goals to the empty set  \label{def:t1}
\item The remaining subgoals are irreducible --- that is, no reduction operators are applicable to the present set of goals \label{def:t2}
\end{enumerate}

\begin{example}[Infinite Reduction]~\label{ex:infinite-reduction}
    Consider the following reduction operator:
    \[
    \rho: (\phi \supset \psi, \Gamma \seq \chi) \mapsto \{ (\phi \supset \psi, \Gamma \seq \phi), (\psi, \phi \supset \psi, \Gamma \seq \chi)  \} 
    \]
    Observe that it essentially corresponds to $\supset$-elimination rule in $\calculus{LJ}$~\cite{Gentzen}. Applying it to the goal $(p \supset p \seq p)$ yields the subgoals $\{ (p \supset p \seq p), (p, p \supset p \seq p)\}$. Applying $\rho$ to either subgoal does not change the list of subgoals. Hence, there is an infinite sequence of reductions,  
    \[
    \{(p \supset p \seq p) \} \overset{\rho}{\longrightarrow} \{ (p \supset p \seq p), (p, p \supset p \seq p)  \} \overset{\rho}{\longrightarrow} \{ (p \supset p \seq p), (p, p \supset p \seq p)  \} \overset{\rho}{\longrightarrow} \ldots
    \]
   This is a typical example of the kind of computations encountered in co-inductive logic programming --- see, for example, Gupta et al.~\cite{gupta2007coinductive,simon2007co}. Note that one can construct a more pathological example than the one given by not only having the computation never terminate, but also having the set of goals increase with each reduction. 
\end{example}

\begin{example}[Termination~\ref{def:t1}] \label{ex:t1}
    Let $\rho_1$ be $\rho$ in Example~\ref{ex:infinite-reduction} and let $\rho_2$ be as follows for any formula $\phi$:
    \[
    \rho_2: (\phi, \Gamma \seq \phi) \mapsto \emptyset
    \]
    Beginning with the goal $(\phi,\phi \supset \psi \seq \psi)$, we have the following sequence:
        \[
    \{(\phi,\phi \supset \psi \seq \psi) \} \overset{\rho_1}{\longrightarrow} \{ (\phi,\phi \supset \psi \seq \phi), (\psi,\phi,\phi \supset \psi  \seq \psi)  \} \overset{\rho_2}{\longrightarrow} \{  (\psi,\phi,\phi \supset \psi  \seq \psi) \} \overset{\rho_2}{\longrightarrow} \emptyset
    \]
    This shows an sequence terminating with \ref{def:t1}.
\end{example}

\begin{example}[Termination~\ref{def:t2}]
    Let $\rho_1$ and $\rho_2$ be as in Example~\ref{ex:t1}. When $\phi \neq \psi$, the goal $(\phi \seq \psi)$ is irreducible using the reduction operators $\{\rho_1, \rho_2\}$. Therefore, it represents a terminating sequence (of one step) of type \ref{def:t2} (not \ref{def:t1}).
\end{example}

Intuitively, T1 means that the putative goal is indeed valid. This determines an important soundness and completeness condition for the semantics of reductive logic. Fix a logic $\mathcal{L}$:
\vspace{1mm}
\begin{enumerate}[align=left]
    \item[\textbf{(Soundness)}] $\setRed$ is \emph{sound} if, for any goal $G \in \setGoals$, if the state $\{G\}$ has a sequence of reductions ending in the empty set, then $G$ is valid in $\mathcal{L}$.
    \item[\textbf{(Completeness)}] $\setRed$ is \emph{complete} if, for any goal $G \in \setGoals$, if $G$ is valid in $\mathcal{L}$, then the state $\{G\}$ has a sequence of reductions ending in the empty set.
\end{enumerate}
\vspace{1mm}
Observe that beginning with a goal $G$ and witnessing an infinite sequence of reduction or a finite one terminating with \ref{def:t2} does not necessarily mean $G$ is \emph{not} valid, only that we have not been able to witness that it is. For these cases to show that the putative goal is invalid, we should need a meta-theorem that says that the sequence of reductions used is itself complete, which is stronger than what is prescribed above. Of course, this relates to the celebrated \emph{negation-as-failure} protocol --- see, for example, Clark~\cite{clark1977negation}.

We regard these conditions as criteria for a semantics of reductive logic. Observe that \ref{def:r1} essentially guarantees that these conditions hold. However, in doing so, it narrows the scope of reduction operators to the point where asking about their validity becomes nonsensical. We return to this in Section~\ref{sec:reductive-logic-2}.

\subsection{The Space of Reductions}

While we have modelled `reduction' through the use of operators and said something about how they compose to form reductive reasoning, we can be much more precise about the space of exploration during reductive reasoning. This space, the space of reductions, is larger than the space of deductions --- that is, the space of objects studied in deductive logic; for example, the class of proofs in a fixed proof-system --- as it also contains `failed' or `incomplete' reductions --- that is, reductions failing to satisfy condition \ref{def:t1}.

To model the space of reductions for a given goal $G\in \setGoals$, we move from considering individual reduction operators to considering \emph{all} possible reductions as different sets of sets of sufficient premisses. Let $\setRed$ denote the set of reduction operators given in \ref{def:r1prime}.  We define a single \emph{destructor} for $\setRed$ representing the collected action of all the reduction operators that apply at any point, 
\[
\partial: \setGoals \to \funcfinPower(\funcfinPower(\setGoals))
\]
with the action
\[
\partial:G \mapsto \{ \rho(G) \mid \rho \in \setRed \}
\]
--- recall that $\setRed$ is finite and each $\rho \in \setRed$ is a $\funcfinPower$-coalgebra. Using $\partial$, one can define the space of reductions for a goal $G$ as a limit of $\partial$ applied to a given goal:

\begin{definition}[Reduction Space]~\label{def:reduction-space} The reduction space $\mathfrak{R}(G)$ for a goal $G \in \setGoals$ is the the greatest tree satisfying the following:
		\begin{enumerate}[label=--]
		\item[--] the root of the tree is a node labelled by $G$,
		\item[--] the root has $|\partial(G)|$ children labelled $\bullet$,
		\item[--] for each $\bullet$, there exists a unique set $\{G_0,\ldots , G_n\} \in \partial(G)$
        \item[--] if $\{G_0,\ldots ,G_n\} \in \partial(G)$ corresponds to $\bullet$, then the children of $\bullet$ are $\mathfrak{R}(G_1), \ldots, \mathfrak{R}(G_n)$.
	\end{enumerate}
\end{definition}

\begin{example}
    In Figure~\ref{fig:reduction-space}, we show the reduction space for $\phi, \phi \supset \phi \seq \phi$. For readability, we have labelled the branches using the generating reduction operators. Here, $\rho_1$ corresponds to the reduction operator in Example~\ref{ex:infinite-reduction} and $\rho_2$ corresponds to the reduction operator in Example~\ref{ex:t1}. Observe that it is an \emph{infinite} tree along the rightmost branch. 
\end{example}

\begin{figure}
     \hrule 
    \[
\xymatrix{
\phi, \phi \supset \phi \seq \phi \ar@{->}[d]^{\rho_2} \ar@{->}[dr]^{\rho_1}  & & & &\\
\bullet \ar@{->}[d] & \bullet \ar@{->}[d] \ar@{->}[drr] & & & \\
\Box    & \phi, \phi \supset \phi \seq \phi \ar@{->}[d]^{\rho_2} \ar@{->}[dr]^{\rho_1}  & & \phi,\phi\supset \phi \seq \phi  \ar@{->}[d]^{\rho_2} \ar@{->}[dr]^{\rho_1} & \\ 
    & \vdots &  \ddots & \vdots & \ddots
}
    \]
    \hrule \vspace{1mm}
    \caption{Reduction Space for $\phi, \phi \supset \phi \seq \phi$} \label{fig:reduction-space}
\end{figure} 

This is the co-inductive version of an \emph{and}/\emph{or}-tree as used in \emph{logic programming} (LP) --- see, for example, \cite{Komendantskaya2010,Komendantskaya2011,Komendantskaya16,Bonchi2015}. This nomenclature arises from the fact that that the $\bullet$ denotes a choice, representing the disjunction of sets of sufficient premisses, and the children of any given or-node must all be redexes, representing a conjunction of sufficient premisses. This suggests the following representation of a reduction sequence:

\begin{definition}[Reduction Tree]
 A reduction tree for $G$ is any tree $\mathcal{R}$ such that $G$ is the root of $\mathcal{A}$ and the immediate subtrees $\mathcal{A}_1 \ldots \mathcal{A}_n$ are the reduction trees for $G_1,\ldots,G_n \in \setGoals$, respectively, where there is $\rho \in \setRed$ and $\rho(G)=\{G_1,\ldots,G_n\}$.
\end{definition}
\begin{example}
    In Figure~\ref{fig:reduction-tree}, we present the reduction sequence in Example~\ref{ex:t1} as a reduction tree in which $\Box$ denotes a node with no sequent.  
\end{example}
\begin{figure}
\hrule
    \[
    \xymatrix{
    & \ar@{->}[dl] (\phi, \phi \supset \psi \seq \psi) \ar@{->}[dr] &  \\
    \ar@{->}[d] (\phi,\phi \supset \psi \seq \phi) &  &  (\psi,\phi,\phi \supset \psi \seq \psi) \ar@{->}[d] \\
    \Box & & \Box
    }
    \]
    \hrule \vspace{1mm}
    \caption{A Reduction Tree for $(\phi, \phi \supset \psi \seq \psi)$}
    \label{fig:reduction-tree}
\end{figure}

One may view the reduction space as the glueing together of all the possible reduction trees using $\bullet$ to denote different choices. Conversely, one may extract them by skipping over $\bullet$-nodes: beginning from the goal, choose \emph{one} $\bullet$ and connect the goal to all the children of that node, then repeat the process for these subgoals --- see Gheorghiu et al~\cite{GDP2023bi-lp}.

Following work by Komendantskaya et al.~\cite{Komendantskaya2010,Komendantskaya2011,Komendantskaya16} (see also Zanasi and Bonchi~\cite{Bonchi2015} and Gheorghiu et al.~\cite{GDP2023bi-lp}), we can model the space of reductions as the cofree comonad $\func{C}$ of 
$\funcfinPower(\funcfinPower(-))$ applied to 
$\setGoals$.\footnote{A \emph{monad} on a category $\mathcal{C}$ is given by an endofunctor $T:\mathcal{C}\rightarrow \mathcal{C}$ 
together with two natural transformations $\eta : 1_\mathcal{C} \rightarrow T$ (`unit'), where 
$1_\mathcal{C}$ denotes the identity functor on $\mathcal{C}$, and $\mu : T^2 \rightarrow T$ 
(`multiplication'), where $T^2$ is the functor $T \circ T : \mathcal{C} \rightarrow \mathcal{C}$ subject 
to the following laws:
\begin{itemize}[label=--]
	\item $\mu \circ T\mu = \mu \circ \mu T : T^3 \rightarrow T$, where $T \mu$ and $\mu T$
	are formed by the induced `horizontal composition' of natural transformations (see \cite{MacLane71}) 
	\item $\mu \circ T \eta = \mu \circ \eta T = 1_T : T \rightarrow T$, where $1_T$ denotes the 
	identity natural transformation form $T$ to $T$.  
\end{itemize}   
A \emph{comonad} is categorical dual on the category $\mathcal{C}^{\text{op}}$, with axioms for 
counit and comultiplication, obtained by reversing the arrows in the definition above. 

\hspace{1em} Generally speaking, `free' means an instance of a structure that satisfies just the 
axioms of its definition, with no additional structure imposed. A \emph{free functor} 
arises as a left adjoint to a forgetful functor and a \emph{cofree functor} arises as 
a right adjoint to a forgetful functor. For example, a left adjoint to the forgetful functor from a given category $\mathcal{C}$ of algebraic structures to the category of sets takes sets $X$ to their corresponding free objects in $\mathcal{C}$. See \cite{nLabFF} for an explanation and examples.} That is, for a given $X$, consider 
the following sequence of sets, where $\alpha$ is any ordinal:
 \[
\begin{cases}
Y_0 &:= X\\
Y_{\alpha+1} &:= \setGoals \times \funcfinPower(\funcfinPower (Y_{\alpha}))
\end{cases}
\]
Each stage of the construction yields a function $\partial_\alpha:X \to Y_\alpha$ defined inductively as follows, where $I$ is the identify function:
 \[
\begin{cases}
\partial_0 &:= I \\
\partial_{\alpha+1} &:=  I \times  (\funcfinPower(\funcfinPower(\partial_\alpha)) \circ \partial)
\end{cases}
\]
This construction converges at the limit ordinal $2\omega$ --- see Worrell~\cite{WORRELL2005184}. We denote the resulting map $\nabla:X \to \func{C}(X)$. This is a standard construction for a cofree comonad. For $X=\setGoals$, this precisely maps goal $G\in \setGoals$ to its space of reductions:

\begin{theorem}\label{thm:coherence}
    For any $G \in \setGoals$, 
    \[
    \nabla(G) = \mathfrak{R}(G)
    \]
\end{theorem}

\begin{proof}
     We proceed by co-induction on the definition of $\mathfrak{R}(G)$ (Definition~\ref{def:reduction-space}). For a proof by co-induction, we use the uniqueness of the greatest fixed point (see Section~\ref{sec:introduction}). That is, we need only show that applying the destructor defining $\mathfrak{R}(G)$ is equivalent to applying the destructor defining $\nabla(G)$, assuming that the results of such applications are equivalent. This is made clear presently. 

Following Definition~\ref{def:reduction-space}, we think of each $\bullet$ as a set of possible subgoals; that is, the elements of $\partial$ when it has been applied to a goal. In this reading, encoding trees a lists, Definition~\ref{def:reduction-space} is expressed as follows:
\[
\mathfrak{R}(G) := \langle G, \{ \{\mathfrak{R}(G_1),\ldots, \mathfrak{R}(G_n)\} \mid \exists \rho \in \setRed: \rho(G)=\{G_1,\ldots G_n\} \} \rangle \tag{$\ast$}
\]
This expresses $\mathfrak{R}$ as a \emph{destructor} for the reduction spaces. 

Let $\setStyle{SUB}(G) := \cup_{\rho \in \setRed} \rho(G)$. The co-induction hypothesis is the following: 
\[
    \text{If $G' \in \setStyle{SUB}(G)$, then $\nabla(G') = \mathfrak{R}(G')$} \tag{co-IH}
\]
The result follows immediately:
\begin{align}
\mathfrak{R}(G) &= \langle G, \{ \{\mathfrak{R}(G_1),\ldots, \mathfrak{R}(G_n)\} \mid \exists \rho \in \setRed: \rho(G)=\{G_1,\ldots G_n\} \} \rangle \tag{$\ast$} \\
&= \langle G, \{ \{\nabla(G_1),\ldots, \nabla(G_n)\} \mid \exists \rho \in \setRed: \rho(G)=\{G_1,\ldots G_n\} \} \rangle \tag{co-IH} \\
&= \nabla(G) \notag
\end{align}
This completes the proof.
\end{proof}

This completes the construction of the space of reductions. 

\subsection{Soundness \& Completeness}

Let $\calculus{L}$ be a deductive system over $\setGoals$. We elide details of $\calculus{L}$ as they do not matter; concrete examples include natural deduction systems, sequent calculi, tableaux systems, and so on.  Fix a logic $\logic{L}$. We have the following soundness and completeness conditions:
\vspace{2mm}
\begin{enumerate}[align=left]
    \item[\textbf{(Soundness)}] $\calculus{L}$ is \emph{sound} for $\logic{L}$ if, for any goal $G \in \setGoals$, if $G$ admits a proof in $\calculus{L}$, then $G$ is valid in $\logic{L}$.   \item[\textbf{(Completeness)}] $\calculus{L}$ is \emph{complete} for $\logic{L}$ if, for any goal $G \in \setGoals$, if $G$ is valid in $\logic{L}$, then $G$ admits a proof in $\calculus{L}$.
\end{enumerate}
\vspace{2mm}
For the remainder of this section, we take a fixed $\calculus{L}$ that is sound and complete for a fixed logic $\logic{L}$.

We can define the correctness of $\setRed$ relative to a calculus $\calculus{L}$ as follows: 
\vspace{2mm}
\begin{enumerate}[align=left]
    \item[\textbf{(Faithfulness)}] $\setRed$ is \emph{faithful} for $\calculus{L}$ if, for any goal $G \in \setGoals$ and any $\rho \in \setRed$, if $\rho(G) = \{G_1,\ldots,G_n\}$, then there is a rule in $\calculus{L}$ from $G_1, \ldots G_n$ to $G$.
    \item[\textbf{(Adequacy)}] $\setRed$ is \emph{adequate} for $\calculus{L}$ if, for any goal $G \in \setGoals$, if there is a rule from $G_1,\ldots,G_n$ to $G$ in $\calculus{L}$, then there is a reduction operator $\rho \in \setRed$ such that $\rho(G) = \{G_1,\ldots,G_n\}$.
\end{enumerate}
\vspace{2mm}
Relative to this set-up, the validity of reduction we automatically have  soundness and completeness of $\setRed$ with respect to $\logic{L}$:

\begin{proposition}[Soundness \& Completeness]
The following both hold:
\begin{enumerate}[label=--]
\item if $\setRed$ is faithful for $\calculus{L}$, then $\setRed$ is sound for $\logic{L}$
\item If $\setRed$ if adequate for $\calculus{L}$, then $\setRed$ is complete for $\logic{L}$. 
\end{enumerate}
\end{proposition}
\begin{proof}
Immediate by induction since reduction operators in $\setRed$ are instances of the inverses of rules in $\calculus{L}$.
\end{proof}

This broadly represents the account of reductive logic used in computer-assisted formal reasoning. A central example is \emph{Automated Theorem Proving} (ATP), which leverages algorithms to emulate the logical reasoning process that human mathematicians use, representing reasoning in many formal systems such as sequent calculi. This includes, for example, proof by induction, case analysis, and so on. It is the heart of automated reasoning tools (see, for example, Portoraro~\cite{sep-reasoning-automated}), including \emph{logic programming} (see, for example, Miller~\cite{miller1989logical,miller1991uniform}), and symbolic AI (see, for example, Bundy~\cite{Bundy1983} and Gilles~\cite{gillies1996artificial}). Of course, ATP applies a \emph{range} of techniques including, \emph{inter alia}, analytic tableaux~\cite{Smullyan1968}, resolution~\cite{Kowalski1979}, the matrix method,~\cite{wallen}, and uniform/focused proof-search~\cite{miller1989logical,miller1991uniform}. In particular, the latter has received much interest because it enables a uniform platform for proof construction as well as other reasoning methodologies (e.g., SAT/SMT solving algorithms such as DPLL~\cite{dpll}). 

Despite its efficacy in the above problem domain, this account of reductive logic is somewhat limited in that, essentially, it demands that reduction operators be exactly the duals of admissible rules, recovering \ref{def:r1}. What is more, it speaks of the correctness of the collection of reduction operators, but not of the validity of reduction operators themselves.  

In the next section,  we give an alternative semantics for reduction operators in which the assertion language of goals need not even be the assertion language of the logic, as long as one has some sense of correspondence between them. 

\section{Reduction Operators II --- Operational Semantics} \label{sec:reductive-logic-2}

In Section~\ref{sec:reductive-logic-1}, we provided a denotational semantics for reduction operators from first principles. This enables the representation of reductive reasoning. In this section, we present a corresponding \emph{operational} semantics, allowing for a more refined analysis that provides a subtle and expressive account of reduction-operator usage.

\subsection{Operational Semantics}

In this section, we introduce \emph{structural operational semantics} (SOS) for reduction and reduction operators. Originally introduced by Plotkin~\cite{plotkin1981structural} to define the behaviour of programming constructs, SOS uses formal rules to describe how each part of a program manipulates a state to yield a new state. It adopts a meaning-as-use approach, where `use' implies `behaviour'.

For our purposes, we focus on one of the most extensively studied rule formats for SOS: the format of \emph{general structural operational semantics} (GSOS) as defined by Bloom et al.~\cite{bloom1995bisimulation}.

We represent a state as a (finite) list of current goals, where $G \in \setGoals$, 
\[
S ::= \Box \mid G::S
\]
Intuitively, $\Box$ denote the empty string. We may write $[G_1, \ldots, G_n]$ to denote the list $G_1::(\ldots::G_n)$.

We have a countable collection of action labels $\mathsf{A}$ and a reduction operator $\rho$ is the transformation of a goal $G$ into a list of subgoals $[G_1,\ldots, G_n]$,
\[
G \overset{\rho}{\longrightarrow} [G_1, \ldots ,G_n] \tag{$\Uparrow_\partial$}
\]
The set of reduction operators $\setRed$ from Definition~\ref{def:r1prime} can be seen as comprising all such rules. Notably, while Section~\ref{sec:reductive-logic-1} used finite sets, we now work with finite strings due to syntactic considerations.

We model the free choice of goal reduction by adding the following for all $\rho \in \mathsf{A}$:
\[
    \infer{ G \longrightarrow [G_1, \ldots ,G_n]}{G \overset{\rho}{\longrightarrow} [G_1, \ldots ,G_n]} \tag{$\uparrow_\rho$}
\]
This behaviour extends to states with more than one element via the following rules:
\[
    \infer{S_1::S_2 \longrightarrow S_1'::S_2'}{S_1  \longrightarrow S_1' &  S_2 \longrightarrow  S_2'} \tag{$\uparrow$}
\]
The operational semantics of reduction comprises the collection of all such rules,
\[
    \mathcal{O} := \ \setRed \cup \{\uparrow\} \cup \{\uparrow_\rho \mid \rho \in \mathsf{A} \}
\]
This system enables us to describe the process of reduction as a `computation' in $\mathcal{O}$.

\begin{example} \label{ex:operational}
    The reduction sequence in Example~\ref{ex:t1} is presented in $\mathcal{O}$ as follows:
    \[
    \infer{(\phi, \phi \supset \psi \seq \psi) \longrightarrow  (\phi, \phi \supset \psi \seq \phi) :: (\psi, \phi, \phi \supset \psi \seq \psi)}{
    (\phi, \phi \supset \psi \seq \psi) \overset{\rho_1}{\longrightarrow}  (\phi, \phi \supset \psi \seq \phi) :: (\psi, \phi, \phi \supset \psi \seq \psi)} 
    \]
    and
    \[
    \infer{(\phi, \phi \supset \psi \seq \phi) :: (\psi, \phi, \phi \supset \psi \seq \psi) \longrightarrow \Box}{
        \infer{(\phi, \phi \supset \psi \seq \phi) \longrightarrow \Box}{(\phi, \phi \supset \psi \seq \phi) \overset{\rho_2}{\longrightarrow} \Box} & \infer{(\psi, \phi, \phi \supset \psi \seq \psi) \longrightarrow \Box}{(\psi, \phi, \phi \supset \psi \seq \psi) \overset{\rho_2}{\longrightarrow} \Box}}
    \]
We discuss combining these to represent a sequence of reductions below.
\end{example}

Following Turi and Plotkin~\cite{turi97}, this GSOS specification can be modelled as a \emph{bialgebra}. Given two functors $S$ and $T$, and a \emph{distributive law} $\lambda:TS \Rightarrow ST$, a bialgebra is a pair of an $T$-algebra $\alpha:TX \to X$ and an $S$-coalgebra $\beta:X \to SX$ over a common carrier (i.e., $X$) satisfying the following \emph{pentagonal} law:
\[
 \beta \circ \alpha = S\alpha \circ \lambda \circ T\beta
\]
 Turi and Plotkin~\cite{turi97} observe that one can construct bialgebras from a GSOS such that the algebra is a denotational model, the coalgebra is an operation model, and the distributive law says that the combination of the two models obeys the rules of the GSOS. 
 
 This approach can be applied immediately to the account of reduction above. Our state space is modelled by $X = \funcList(\setGoals)$. The heuristics are as follows:
\begin{enumerate}
    \item[--] the internal structure of states is modelled by an algebra $\alpha:\funcList (X) \to X$ --- thus, by ($\uparrow$), a state $S_1::S_2$ is regarded as the concatenation of the lists of states $S_1$ and $S_2$,
    \item[--]  the behaviour (i.e., reduction) is modelled by a coalgebra $\beta: X \to \funcfinPower(\funcList(X))$; that is, by $\uparrow_\rho$, a state is taken to the set of next possible states,
    \[
   S \mapsto \{ S' \in \funcList(\setGoals) \mid \exists \rho \in \mathsf{A}: S \overset{\rho}{\longrightarrow} S'\}
    \]
\item[--] there is a distributive law $\lambda:\funcList(\funcfinPower) \Rightarrow \funcfinPower(\funcList)$ ensuring coherence,
\[
[X_1,\ldots,X_n] \mapsto \{ [x_1,\ldots,x_n] \mid \mbox{$x_i \in X_i$ for $i=1,\ldots n$}\}
\]
\end{enumerate}
In the behaviour functor for $\beta$ --- that is, $\funcfinPower(\funcList(-))$ --- the internal functor $\funcList$ represents the structure of states and is conjunctive, because all the goals are potential, while the external functor $\funcfinPower$ is disjunctive, because only one collection of sufficient premisses needs to be checked. This reading justifies the particular distributive law used. For more technical details, see Bonchi and Zanasi~\cite{Bonchi2015}.

This operational semantics can be easily extended to express \emph{large} steps of reduction (rather than just small ones). For example, we take the reflexive and transitive closure:
\[
S \longrightarrow S
\qquad
\infer{S \longrightarrow S''}{S \longrightarrow S' & S' \longrightarrow S''}
\]
Now, $G \longrightarrow \Box$ means that there is some sequence of reduction terminating with an empty list of remaining subgoals (\ref{def:t1}). 

\begin{example}
In the large-step operational semantics,  $(\phi, \phi \supset \psi \seq \psi) \longrightarrow \Box$. This is witnessed by the computation
  \[
  \infer{(\phi, \phi \supset \psi \seq \psi) \longrightarrow \Box}{
            \deduce{(\phi, \phi \supset \psi \seq \psi) \longrightarrow  (\phi, \phi \supset \psi \seq \phi) :: (\psi, \phi, \phi \supset \psi \seq \psi)}{\vdots}
            &
            \deduce{(\phi, \phi \supset \psi \seq \phi) :: (\psi, \phi, \phi \supset \psi \seq \psi)
            \longrightarrow \Box}{\vdots}
    }
    \]
in which both the left and right branches continue exactly as in Example~\ref{ex:operational}.
\end{example}

A `composition' can be seen as any (partial) mapping that from a collection of reduction operators maps a new reduction operator. Formally, let $\setStyle{R}$ denote the set of all reduction mappings: $\rho \in \setStyle{R}$ if $\rho:\setGoals \to \funcList(\setGoals)$. A composition is a (partial) mapping $\circ:\setStyle{R}^n \partialto \setStyle{R}$. Intuitively, $\circ$ is valid if it preserves validity: for any $\rho_1,\ldots,\rho_n \in \setStyle{R}$, if $\rho_1,\ldots,\rho_n$ are all valid, then $\circ(\rho_1,\ldots,\rho_n)$ is valid.

This approach, as articulated by Milner~\cite{milner1984tactics}, allows us to demonstrate the validity of a proposed goal in terms of individual steps. As Milner~\cite{milner1984tactics} notes:
\begin{quote}
    `Here it is a matter of taste whether the human prover wishes to see this performance done by the machine, in all its frequently repulsive detail, or wishes only to see the highlights, or is merely content to let the machine announce the result (a theorem!).'
\end{quote}

Various kinds of composition can be given operational semantics. For example, we can extend the language of actions with symbols $\seqcomp$ , $+$ , and $(-)^\ast$: 
\[
a ::= \rho \in \mathsf{A} \mid a\seqcomp a \mid a+a \mid a^\ast
\]
By adding appropriate rules to $\mathcal{O}$, we define $\seqcomp$ as `sequential composition of reduction operators', $+$ as `non-deterministic choice of reduction operators', and $\ast$ as `arbitrary number of applications' by the following rules:
\[
\begin{array}{c}
\infer[\uparrow_{\seqcomp}]{S \overset{a\seqcomp b}{\longrightarrow} S''}{S  \overset{a}{\longrightarrow} S' &  S' \overset{b}{\longrightarrow}  S''} \qquad 
\infer[\uparrow_{+}^1]{S \overset{a+b}{\longrightarrow} S'}{S  \overset{a}{\longrightarrow} S'} 
\quad 
\infer[\uparrow_{+}^2]{S \overset{a+b}{\longrightarrow} S'}{S  \overset{b}{\longrightarrow} S'} \\
\infer[\uparrow_\ast^1]{S \overset{a^\ast}{\longrightarrow} S''}{S \overset{a}{\longrightarrow} S' & S' \overset{a^\ast}{\longrightarrow} S''}
  \qquad 
\infer[\uparrow_\ast^2]{S \overset{a^\ast}{\longrightarrow} S}{}
\end{array}
\]
This language is a relatively refined way of expressing sequences of reductions in terms of process algebra --- see, for example, Milner~\cite{milner1983calculi,milner1989communication} and Bergstra and Klop~\cite{bergstra1984algebra}.

We have thus provided an operational semantics for reduction, but what determines its correctness with respect to a formal system? We can apply the soundness and completeness criteria from Section~\ref{sec:reductive-logic-1}, suitably translated. 
\vspace{2mm}
\begin{enumerate}[align=left]
\item[\textbf{(Soundness)}] $\setRed$ is \emph{sound} for $\logic{L}$ if, for any goal $G \in \setGoals$, if $G \longrightarrow \Box$, then $G$ is valid in $\logic{L}$.
\item[\textbf{(Completeness)}] $\setRed$ is \emph{complete} for $\logic{L}$ if, for any goal $G \in \setGoals$, if $G$ is valid in $\logic{L}$, then $G \longrightarrow \Box$.
\end{enumerate}
\vspace{2mm}


This operational semantics satisfactorily explains what reduction operators are. However, understanding what makes them \emph{valid} is crucial. We can assert that reduction operators are valid when there exists a corresponding admissible rule for the logic in which we wish to reason, following the approach outlined in Section~\ref{sec:reductive-logic-1}. In the next section, we refine this validity condition by relating it to \emph{the theory of tactical proof}, as introduced by Milner~\cite{milner1984tactics}

\subsection{The Theory of Tactical Proof} \label{sec:inferentialism}

A foundational framework that supports the mechanization of reductive logic is the theory of \emph{tactical proof}. It systematically underpins the proof-assistants mentioned in Section~\ref{sec:introduction}. The centrality of tactical proof to computer assisted formal reasoning cannot be understated --- see, for example, Gordon~\cite{gordon2015tactics}. Its efficacy is that it allows users to articulate insight into reasoning methods and delegate routine but possibly error-prone work to machines. The theory clarifies how concepts such as `goal', `strategy', `achievement', `failure', etc., interrelate during reductive reasoning. In this section, we observe that tactical proof can be viewed as a semantics of reduction operators. 

To simplify the operational semantics, we define a set of primary entities called goals ($\setGoals$) and reduction operators given by mappings:
\[
\rho:\setGoals \partialto \funcList(\setGoals)
\]
Milner~\cite{milner1984tactics} extends this set-up by introducing another set of primary entities called \emph{events} ($\setEvents$) and an \emph{achievement} relation $\propto$, which specifies how events witness goals:
\[
\propto \subseteq \setGoals \times \setEvents
\]
This allows us to express that certain events satisfy certain goals.

\begin{example} \label{ex:event}
    Consider $G$ from Example~\ref{ex:Alice-Bob-1}, asserting that Alice and Bob meet before noon on Saturday. Let $E$ be the event of `Alice and Bob meeting under the clock at Waterloo station at 11:53 on Saturday.' Here, $E$ \emph{achieves} $G$ as it satisfies the designated description of the goal.
\end{example}

We now ask: \emph{What makes a reduction operator invalid}? Intuitively, events that satisfy subgoals should establish events that satisfy the original goal. As noted by Milner~\cite{milner1984tactics}, in Example~\ref{ex:Alice-Bob-1}, if `noon' were replaced by `evening' in both $G_1$ and $G_2$, then $\rho$ should be deemed invalid. Therefore, a reduction operator is \emph{valid} when any events achieving subgoals can be justified to suffice for an event achieving the goal. Milner~\cite{milner1984tactics} refers to this justification as a \emph{procedure}:
\[
\pi:\funcList(\setEvents) \partialto \setEvents
\]
Note that, like reduction operators, procedures are partial functions.

\begin{example}
    To complement $G_1$ and $G_2$ from Example~\ref{ex:Alice-Bob-1}, asserting that Alice and Bob arrive at Waterloo before noon on Saturday, consider the events:
    \begin{enumerate}[label=--]
        \item $E_1$: \text{Alice arrives at Waterloo Station at 11:57 on Saturday}
        \item $E_2$: \text{Bob arrives at Waterloo Station at 11:53 on Saturday}
    \end{enumerate}
    Here, $E_1$ achieves $G_1$ and $E_2$ achieves $G_2$. Moreover, the event $E$ in Example~\ref{ex:event} arises from $E_1$ and $E_2$ via the \emph{wait} procedure, which requires Alice and Bob to wait for each other.
\end{example}

Milner~\cite{milner1984tactics} defines a \emph{tactic} as a reduction operator that also provides the procedure justifying the subgoals. This formalizes the validity condition described above. Let $\Pi$ denote the set of procedures, a subset of all functions $\pi:\funcList(\setEvents) \partialto \setEvents$.

\begin{definition}[Tactic]
    A \emph{tactic} is a partial mapping $\tau:\setGoals \to \funcList(\setGoals) \times \Pi$.
\end{definition}

\begin{definition}[Validity for Tactics]
    A tactic $\tau$ is \emph{valid} if, for any $G \in \setGoals$, whenever $\tau:G \mapsto \langle [G_1,\ldots,G_n], \pi \rangle$ and $E_1,\ldots, E_n \in \setEvents$ are such that $G_i\propto E_i$ for $i=1\ldots n$, then $G \propto \pi(E_1,\ldots,E_n)$.
\end{definition}

A tactic is a reduction operator that upon application to a goal produces a list of subgoals as well as a witness that the reduction is valid (i.e., the procedure). This provides a robust account of reductive reasoning widely applied in mechanized reasoning (see Gordon~\cite{gordon2015tactics}). However, we require only that procedures \emph{exist} to account of the validity of the reduction itself. Thus:

\begin{definition}[Validity for Reduction Operators]
    A reduction operator $\rho$ is valid if, for any $G \in \setGoals$, whenever $\rho:G \mapsto [G_1,\ldots,G_n]$ and $E_1,\ldots, E_n \in \setEvents$ are such that $G_i\propto E_i$ for $i=1\ldots n$, then there exists $\pi \in \propto$ such that $G \propto \pi(E_1,\ldots,E_n)$.
\end{definition}

This formulation of validity provides a satisfactory account of reduction and reduction operators as it separates the action of `reduction' from the validity of such an action. 

\begin{example}[Milner's Proof-search Machine~\cite{milner1984tactics}]\label{ex:tacticsofcalculus}
    Let $\proves$ be a \emph{Tarskian}~\cite{tarski1936concept} consequence relations --- in particular, it is monotonic. The set-up is as follows:
    \begin{itemize}[label=--]
        \item the set of \emph{goals} is defined to be the set of sequents $\Gamma \seq \phi$, in which $\Gamma$ is a list of formulas and $\phi$ is a formula
        \item the set of \emph{events} is defined to be the set of consequences --- that is, sequents $\Delta \seq \psi$ such that $\Delta \proves \psi$
        \item the \emph{achievement} relation $\propto$ is defined as follows:
        \[
        (\Gamma \seq \phi) 
 \propto (\Delta \proves \psi) \text{ if and only if } \phi = \psi \text{ and } \Delta \sqsubseteq \Gamma,
        \]
        where $\Delta \sqsubseteq \Gamma$ means that every element in $\Delta$ is also in $\Gamma$.
    \end{itemize}
    In this set-up, a tactic or reduction operator is valid if it has a procedure corresponding to an admissible rule for the logic.
\end{example}

Observe that the semantics here is one based on `proof' rather than `truth'. That is, according the notion of validity thus provided, what makes a reduction (or sequence of reductions) valid is, ultimately, that it represents a valid forwards inference. This is not a trivial observation as it stands in contrast to the dominant notion of semantics within logic; namely, \emph{model-theoretic semantics} (M-tS).

In M-tS, validity is set up using an abstract notion of \emph{model} $\mathfrak{M}$, satisfying a 
 range of conditions, together with a 
 satisfaction relation $\vDash$ between 
 models and formulas.  A formula $\phi$ is \emph{true} in a model $\mathfrak{M}$ if the model satisfies the formula; that is, $\mathfrak{M} \vDash \phi$. A formula is $\phi$ \emph{valid} iff it is true in all models. This is the paradigm in which one has, for example, the possible-worlds semantics for IL by Beth~\cite{Beth1955} and Kripke~\cite{kripke1965semantical}. 
 
Following Schroeder-Heister~\cite{SEP-IPL}, the paradigm of meaning based on `proof'  is known as \emph{proof-theoretic semantics} (P-tS)~\cite{SEP-PtS,wansing2000idea,francez2015proof}. Importantly, here `proof' does not mean proof in a fixed formal system, but rather constructions based on some \emph{a priori} notion of inference. In the semantics for reductive logic above, this notion of inference is the notion of \emph{procedure}. While P-tS has historically received little attention, most of which restricted to IL (cf. Schroeder-Heister~\cite{SEP-PtS,Schroeder2006validity}), the P-tS of various logical systems has recently seen rapid development; see, for example,  Makinson~\cite{makinson2014inferential}, Nascimento et al.~\cite{nascimento2023ecumenical,stafford2023,nascimentothesis}, Piecha et al.~\cite{Piecha2015failure,Piecha2016completeness,Piecha2019incompleteness}, Pym et al.~\cite{ggp2023imll,ggp2023bi,ggp2024practice,Eckhardt,pym2024categorical}, Sandqvist~\cite{Sandqvist2005inferentialist,Sandqvist2009CL,Sandqvist2015IL}, Schroeder-Heister~\cite{Schroeder2006validity,Schroeder2007modelvsproof,Schroeder1984natural}, and Stafford~\cite{Stafford2021}. 

In \cite{GPTactical}, P-tS is seen to give a semantics for tactical proof. In this paper, tactical proof arises as a special instance of a P-tS of reductive logic. The two accounts are mutually coherent. 
Accordingly, we observe that P-tS as a paradigm of meaning in logic is both historically and practically important for the understanding of logic as a mathematics of reasoning, though rarely expressed as such.

\section{Control R\'egimes} \label{sec:control}
While in the space of proofs, a rule takes a list of premisses and derives a single conclusion, navigating the space of reductions involves many choices that impact the validity of the resulting construction. We refer to all the strategic aspects of managing search in reductive logic as `control'.

\begin{example}\label{ex:multiplicative-and}
Consider the following introduction rule for a multiplicative conjunction $\ast$ as found in substructural logics such as \emph{linear logic}~\cite{Girard1987} and the \emph{logic of bunched implications}~\cite{o1999logic}:
\[
\infer{\Gamma_1,\Gamma_2 \seq \phi_1 \ast \phi_2}{\Gamma_1 \seq \phi_1 & \Gamma_2 \seq \phi_2}
\]
It may be read as a reduction operator as follows:
\[
    \frac{\Gamma_1 \seq \phi_1 \quad \Gamma_2 \seq \phi_2}{\Gamma \seq \phi_1 \ast \phi_2} 
    \,\, (\Gamma = \Gamma_1 , \Gamma_2) \quad \big \Uparrow 
\]
Here, it is necessary to divide $\Gamma$ into two parts, $\Gamma_1$ and $\Gamma_2$. The reduction operator doesn't specify which splitting to use; thus, in applying it, one must \emph{choose} a particular one.  
\end{example}

The control problems may actually be summarized by just two kinds of choices:
\begin{enumerate}[label=\textbf{C\arabic*}]
    \item The choices, which are mutually dependent, of which reduction operator to apply and to which premiss \label{def:c1}
    \item The choice to \emph{backtrack} — that is, to return to a previous point in the search where a choice was made and make a different choice \label{def:c2}
\end{enumerate}

A \emph{control r\'egime} is a specification governing the use of the reduction operators; it is the policy to which we refer when considering these choices.

The application of a control r\'egime delivers a \emph{proof-search} algorithm. Instantiating Kowalski's~\cite{Kowalski1979} celebrated maxim, `$\emph{Algorithm} = \emph{Logic} + \emph{Control}$', in the context of reductive logic, Pym and Ritter~\cite{pym2004reductive} gave the following slogan: 
\vspace{2mm}
\begin{center}
    \textbf{Proof-search} = \textbf{Reductive Logic} + \textbf{Control} 
\end{center}
\vspace{2mm}
It is control that determines the efficacy of proof-search procedures: some procedures will be complete, some not; some will affect the shape of proofs being found, and some will affect the complexity of the procedure. The more control structure provided, the more work that is delegated: mechanical problem-solving begets algorithmic theorem-proving techniques, which beget a programming language paradigm known as logic programming (LP). A well-known example of this is provided by implementations of the logic programming language Prolog, in which the typical strategy is `leftmost literal first' in SLD resolution --- see, for example, Kowalski and Kuehner~\cite{KOWALSKI1971227} and Plaisted and Zhu~\cite{Plaisted1997}. So central are these problems to the use of reductive logic that control is surely a first-class citizen.

How can we represent and reason about control r\'egimes? We have seen that it is possible to extend the operational semantics for reduction with a process calculus, allowing for a flexible way to express control flow during reduction. This setup is more expressive than what can be achieved through the composition of reduction operators (or tactics): while sequential composition ($\seqcomp$) can be modelled by a composition of reduction operators, non-deterministic choice ($+$) cannot.

Arguably, \ref{def:c1} and \ref{def:c2} are not aspects of reduction operators but exist at a higher perspective, as they pertain to reductive reasoning rather than the steps comprising it. This is most evident in  \ref{def:c2}: to backtrack, we need a notion of the sequence of reduction steps taken so far, not just the reduction operators available. Therefore, our notion of state when modelling control cannot be only the list of goals but rather a representation of the explored section of the reduction space.

Related to these control problems is the use of information from one part of the reduction to inform other parts. For example, if we find a successful reduction of a certain subgoal, we should like to reuse that information if we encounter it again during the reduction process. More interesting is how such information may be used dynamically:

\begin{example}
    One way to handle Example~\ref{ex:multiplicative-and} is the so-called \emph{input/output} method introduced by Hodas and Miller~\cite{Hodas1994}, and illustrated in Figure~\ref{fig:in/out}.
    
\begin{figure}[t]
\hrule
\vspace{3mm}
    \centering
    \includegraphics[scale=0.35]{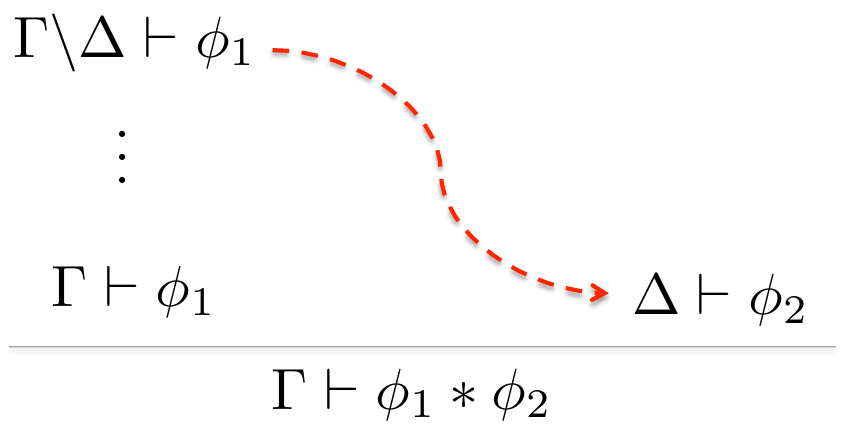} 
    \raisebox{6mm}{\big \Uparrow}
\vspace{3mm}
\hrule \vspace{1mm}
\caption{The Input/Output Method}
\label{fig:in/out}
\end{figure}
Roughly, first, all of $\Gamma$ is sent to the branch for $\Gamma_1$, which is then calculated according to the reduction on the branch. What is left of $\Gamma$ is then sent to the next branch, and the calculation continues.
\end{example}

Gheorghiu and Pym~\cite{GP2023acs} have initiated a framework for studying how information that is acquired during reduction can be used in this way. See also Harland and Pym~\cite{Harland1997,Pym2003}.

We leave these questions for future work. While important to reductive logic and its applications, they are beyond the semantic foundations of reduction operators, demanding a broader perspective of reductive reasoning.

\section{Conclusion} \label{sec:conclusion}
While deductive logic provides a formal foundation for defining proofs through `forwards' inference, the dual consideration of how one constructs proofs using `backwards' inference is essential for understanding the use of logic when reasoning about a putative conclusion. This is reductive logic. It has profound implications across various fields, including program verification, natural language processing, and knowledge representation. By embracing reductive logic, we can not only enhance our understanding of formal reasoning but also develop more effective tools and methodologies for tackling complex problem-solving tasks.

This paper has explored the semantic foundations of reduction operators. They are the reductive logic analogues of rules in deductive logic. We have provided two semantic foundations for them: one denotational and one operational. While the former is conceptually simpler, the operational one permits a more refined account of important aspects of reduction operators, such as the representation of search strategies and their validity. Note that we may regard the operational semantics as an \emph{inferential} semantics of reductive logic in the sense of Brandom~\cite{Brandom2000}.

Essentially, a reduction operator is modelled by a partial function $\rho: \setGoals \partialto \mathcal{F}(\setGoals)$ in which $\setGoals$ is the set of assertions over which we do the reductive reasoning, and $\mathcal{F}(\setGoals)$ denotes collections of such assertions (e.g., finite sets or lists). Relative to this understanding, we can express validity relative to a logic in two different ways.

First, a collection of reduction operators is \emph{valid} (with respect to a logic) if co-recursively reducing a goal eventually produces an empty list just in case the goal is valid in the logic of interest. While intuitive and useful, this does not improve on the extant semantics of reduction by Pym and Ritter~\cite{pym2004reductive}; indeed, it is perhaps \emph{less} informative as, unlike the earlier work, it does not explain in the case of failed searches what extension of the logic is required to render them valid. It also does not quite answer the question: we desire an account of validity that expresses the correctness of the reduction operators themselves (as opposed to their constructions).

Second, a given reduction operator is valid if there corresponds to it a pre-approved \emph{process} such that whenever the premisses are achieved, the conclusion is also achieved. The notion of achievement here is flexible, enabling many diverse situations to be encapsulated. A process corresponds to an accepted forwards inference. This version of validity is essentially Milner's theory of tactical proof~\cite{milner1984tactics}. Thus, we regard tactical proof as `proof-theoretic semantics'~\cite{SEP-PtS} for reductive logic, and this explains the effectiveness of the framework for proof-assistants.

As we have remarked in Section~\ref{sec:introduction}, this paper should be seen as a companion to \cite{GPTactical}. Whereas in \cite{GPTactical} proof-theoretic semantics is used to provide a semantics for tactical proof, this paper provides a general foundation for reductive reasoning in terms of proof-theoretic semantics, establishes direct connections with foundational abstract work in the semantics of computation, and obtains tactical proof as a significant instance.

The scope of this work is the semantic foundations of reduction operators. It has included the semantic foundations of the things constructed by reduction operators. Future work includes giving the semantic foundations for the method by which those constructions (i.e., reductions) are built using reduction operators. This involves, among other things, expressing control regimes, determining the strategy used when exploring the space of reductions, and expressing how one may leverage information about explored portions of the space of reductions to inform future choices. Bundy~\cite{bundy1998proof} has introduced the notion of a \emph{proof plan} as a meta-theoretic tool to represent and reason about the pattern of proof-search and the structure of proofs. Future work also includes developing a semantic account of proof plans as an extension of this paper.


\bibliography{bib}

\begin{thebibliography}{108}
\providecommand{\natexlab}[1]{#1}
\providecommand{\url}[1]{{#1}}
\providecommand{\urlprefix}{URL }
\providecommand{\doi}[1]{\url{https://doi.org/#1}}
\providecommand{\eprint}[2][]{\url{#2}}
 \bibcommenthead

\bibitem[{van Atten(2022)}]{SEP-IPL}
van Atten M (2022) {The Development of Intuitionistic Logic}. The {Stanford}
  Encyclopedia of Philosophy

\bibitem[{Barendregt(1991)}]{Barendregt1991}
Barendregt H (1991) {Introduction to Generalized Type Systems}. Journal of
  Functional Programming 1(2):125--154

\bibitem[{Barendregt et~al(2013)Barendregt, Dekkers, and Statman}]{BDS2013}
Barendregt H, Dekkers W, Statman R (2013) Lambda Calculus with Types.
  Perspectives in Logic, Cambridge University Press

\bibitem[{Barendregt(1993)}]{Barendregt1993}
Barendregt HP (1993) Lambda calculi with types, Oxford University Press, Inc.,
  USA, p 117–309

\bibitem[{Barwise and Moss(1996)}]{Barwise1996}
Barwise J, Moss L (1996) {Vicious Circles: On the Mathematics of
  Non-Wellfounded Phenomena}. Cambridge University Press

\bibitem[{Beaney and Raysmith(2024)}]{sep-analysis}
Beaney M, Raysmith T (2024) {Analysis}. In: Zalta EN, Nodelman U (eds) The
  {Stanford} Encyclopedia of Philosophy, {F}all 2024 edn. Metaphysics Research
  Lab, Stanford University

\bibitem[{Bergstra and Klop(1984)}]{bergstra1984algebra}
Bergstra JA, Klop JW (1984) {The algebra of recursively defined processes and
  the algebra of regular processes}. International Colloquium on Automata,
  Languages, and Programming pp 82--94

\bibitem[{Bertot and Cast\'{e}ran(2004)}]{Bertot2004}
Bertot Y, Cast\'{e}ran P (2004) Interactive Theorem Proving and Program
  Development. Springer, \doi{10.1007/978-3-662-07964-5}

\bibitem[{Beth(1955)}]{Beth1955}
Beth EW (1955) {Semantic Construction of Intuitionistic Logic}. Indagationes
  Mathematicae 17(4):pp. 327--338. \doi{10.1016/S1385-7258(55)50014-6},
  \urlprefix\url{https://doi.org/10.1016/S1385-7258(55)50014-6}

\bibitem[{Bloom et~al(1995)Bloom, Istrail, and Meyer}]{bloom1995bisimulation}
Bloom B, Istrail S, Meyer AR (1995) {Bisimulation can't be traced}. Journal of
  the ACM 42(1):232--268

\bibitem[{Bonchi and Zanasi(2015)}]{Bonchi2015}
Bonchi F, Zanasi F (2015) {Bialgebraic Semantics for Logic Programming}.
  {Logical Methods in Computer Science} 11(1). \doi{10.2168/LMCS-11(1:14)2015}

\bibitem[{Brandom(2000)}]{Brandom2000}
Brandom R (2000) {Articulating Reasons: An Introduction to Inferentialism}.
  Harvard University Press

\bibitem[{Brouwer(1913)}]{brouwer1913intuitionism}
Brouwer LEJ (1913) {Intuitionism and Formalism}. Bulletin of the American
  Mathematical Society 20(2):81--96

\bibitem[{Bundy(1985)}]{Bundy1983}
Bundy A (1985) {The Computer Modelling of Mathematical Reasoning}. Academic
  Press Professional, Inc.

\bibitem[{Bundy(1998)}]{bundy1998proof}
Bundy A (1998) Proof planning. Tech. rep., University of Edinburgh, Department
  of Artificial Intelligence

\bibitem[{Cartmell(1978)}]{cartmell}
Cartmell J (1978) {Generalised Algebraic Theories and Contextual Categories}.
  PhD thesis, Oxford University

\bibitem[{Cheng(2022)}]{cheng2022joy}
Cheng E (2022) {The Joy of Abstraction. An exploration of math, category
  theory, and life}. Cambridge University Press

\bibitem[{Clark(1977)}]{clark1977negation}
Clark KL (1977) {Negation as Failure}. Logic and Data Bases pp 293--322

\bibitem[{{Coq Team}(2024)}]{Inria}
{Coq Team} (2024) {The Coq Proof Assistant}. \url{https://coq.inria.fr/}.
  Access\-ed July 2024.

\bibitem[{van Dalen(2012)}]{vanDalen}
van Dalen D (2012) {Logic and Structure}. Universitext, Springer

\bibitem[{Eckhardt and Pym(2024)}]{Eckhardt}
Eckhardt T, Pym DJ (2024) {Proof-theoretic Semantics for Modal Logics}. Logic
  Journal of the IGPL In Press - arXiv:2401.13597

\bibitem[{Farooque et~al(2013)Farooque, Graham-Lengrand, and Mahboubi}]{dpll}
Farooque M, Graham-Lengrand S, Mahboubi A (2013) A Bisimulation between DPLL(T)
  and a Proof-search Strategy for the Focused Sequent Calculus, Association for
  Computing Machinery, p 3–14. \doi{10.1145/2503887.2503892}

\bibitem[{Fitch(1952)}]{fitch1952symbolic}
Fitch FB (1952) {Symbolic Logic: An Introduction}. Ronald Press Co

\bibitem[{Francez(2015)}]{francez2015proof}
Francez N (2015) {Proof-theoretic Semantics}. College Publications

\bibitem[{{Frank Pfenning, Carsten Sch{\"{u}}rmann, Brigitte Pientka, Roberta
  Vigo, and Kevin Watkins}(2024)}]{Twelf}
{Frank Pfenning, Carsten Sch{\"{u}}rmann, Brigitte Pientka, Roberta Vigo, and
  Kevin Watkins} (2024) {The Twelf Project}.
  \url{http://twelf.org/wiki/Main_Page}. Access\-ed July 2024

\bibitem[{Friedman(1975)}]{Friedman1975}
Friedman H (1975) Some systems of second order arithmetic and their use. In:
  Proceedings of the International Congress of Mathematicians, vol~1. Canadian
  Mathematical Congress, pp 235--242

\bibitem[{Gheorghiu and Pym(2023{\natexlab{a}})}]{GPTactical}
Gheorghiu A, Pym DJ (2023{\natexlab{a}}) {Proof-theoretic Semantics and
  Tactical Proof}. arXiv:230102302 Accessed July 2024

\bibitem[{Gheorghiu and Pym(2023{\natexlab{b}})}]{GP2023acs}
Gheorghiu AV, Pym DJ (2023{\natexlab{b}}) {Defining Logical Systems via
  Algebraic Constraints on Proofs}. Journal of Logic and Computation
  \doi{10.1093/logcom/exad065}

\bibitem[{Gheorghiu and Pym(2023{\natexlab{c}})}]{GP2023semantical}
Gheorghiu AV, Pym DJ (2023{\natexlab{c}}) {Semantical Analysis of the Logic of
  Bunched Implications}. Studia Logica 111(4):525--571

\bibitem[{Gheorghiu et~al(2023{\natexlab{a}})Gheorghiu, Docherty, and
  Pym}]{GDP2023bi-lp}
Gheorghiu AV, Docherty S, Pym DJ (2023{\natexlab{a}}) {Reductive Logic,
  Proof-Search, and Coalgebra: A Perspective from Resource Semantics}. Samson
  Abramsky on Logic and Structure in Computer Science and Beyond pp 833--875

\bibitem[{Gheorghiu et~al(2023{\natexlab{b}})Gheorghiu, Gu, and
  Pym.}]{ggp2023imll}
Gheorghiu AV, Gu T, Pym. DJ (2023{\natexlab{b}}) {Proof-theoretic Semantics for
  Intuitionistic Multiplicative Linear Logic}. In: Automated Reasoning with
  Analytic Tableaux and Related Methods --- TABLEAUX. Springer, pp 367--385

\bibitem[{Gheorghiu et~al(2024)Gheorghiu, Gu, and Pym}]{ggp2024practice}
Gheorghiu AV, Gu T, Pym DJ (2024) {A Note on the Practice of Logical
  Inferentialism}. In: 2nd Logic and Philosophy conference, in Press ---
  arXiv:2403.10546

\bibitem[{Gillies(1996)}]{gillies1996artificial}
Gillies D (1996) {Artificial Intelligence and Scientific Method}. Oxford
  University Press

\bibitem[{Girard(1987)}]{Girard1987}
Girard JY (1987) {Linear Logic}. Theoretical Computer Science 50(1):1--101

\bibitem[{Gordon and Melham(1993)}]{Gordon1993}
Gordon MJ, Melham TF (1993) Introduction to HOL. A theorem proving environment
  for higher order logic. Cambridge University Press

\bibitem[{Gordon et~al(1979)Gordon, Wadsworth, and Milner}]{Gordon1979}
Gordon MJ, Wadsworth CP, Milner R (1979) Edinburgh LCF. A mechanised logic of
  computation. Springer, \doi{10.1007/3-540-09724-4}

\bibitem[{Gordon(2015)}]{gordon2015tactics}
Gordon MJC (2015) Tactics for mechanized reasoning: a commentary on milner
  (1984)`the use of machines to assist in rigorous proof'. Philosophical
  Transactions of the Royal Society A: Mathematical, Physical and Engineering
  Sciences 373(2039)

\bibitem[{Gu et~al(2023)Gu, Gheorghiu, and Pym}]{ggp2023bi}
Gu T, Gheorghiu AV, Pym DJ (2023) {Proof-theoretic Semantics for the Logic of
  Bunched Implications}. arXiv:231116719 Accessed February 2024

\bibitem[{Gupta et~al(2007)Gupta, Bansal, Min, Simon, and
  Mallya}]{gupta2007coinductive}
Gupta G, Bansal A, Min R, et~al (2007) {Coinductive Logic Programming and its
  Applications}. In: International Conference Logic Programming --- ICLP.
  Springer, pp 27--44

\bibitem[{Harland and Pym(1997)}]{Harland1997}
Harland J, Pym DJ (1997) {Resource-distribution via Boolean Constraints}. In:
  Automated Deduction --- CADE. Springer, pp 222--236

\bibitem[{Harland and Pym(2003)}]{Pym2003}
Harland J, Pym DJ (2003) {Resource-distribution via Boolean Constraints}. ACM
  Transactions on Computational Logic 4(1):56--90

\bibitem[{Heyting(1989)}]{heyting1966intuitionism}
Heyting A (1989) {Intuitionism: An Introduction}. Cambridge University Press

\bibitem[{Hodas and Miller(1994)}]{Hodas1994}
Hodas JS, Miller D (1994) {Logic Programming in a Fragment of Intuitionistic
  Linear Logic}. Information and Computation 110(2):327--365

\bibitem[{Hofmann(1997)}]{Hofmann1997}
Hofmann M (1997) {Syntax and Semantics of Dependent Types}. Semantics and
  Logics of Computation p 79–130. \doi{10.1017/CBO9780511526619.004}

\bibitem[{Howard(1980)}]{howard1980formulae}
Howard WA (1980) {The Formulae-as-Types Notion of Construction}. To H B Curry:
  Essays on Combinatory Logic, Lambda Calculus and Formalism 44:479--490

\bibitem[{Jacobs(1991)}]{Jacobs}
Jacobs B (1991) {Categorical Type Theory}. PhD thesis, The University of
  Nijmegen

\bibitem[{Jacobs(2017)}]{jacobs2017introduction}
Jacobs B (2017) {Introduction to Coalgebra}, vol~59. Cambridge University Press

\bibitem[{Kleene(1961)}]{kleene2013mathematical}
Kleene SC (1961) {Mathematical Logic}. Wiley and Sons

\bibitem[{Knaster and Tarksi(1928)}]{Knaster1928}
Knaster B, Tarksi A (1928) {Un Th\'{e}or\`{e}me sur les Fonctions d'Ensembles}.
  Annales de la Societe Polonaise de Mathematique 6:133--134

\bibitem[{Kolmogorov(1932)}]{kolmogorov}
Kolmogorov A (1932) {Zur Deutung der Intuitionistischen Logik}. Mathematische
  Zeitschift 35

\bibitem[{Komendantskaya and Power(2011)}]{Komendantskaya2011}
Komendantskaya E, Power J (2011) {Coalgebraic Semantics for Derivations in
  Logic Programming}. International Conference on Algebra and Coalgebra in
  Computer Science pp 268--282

\bibitem[{Komendantskaya et~al(2011)Komendantskaya, McCusker, and
  Power}]{Komendantskaya2010}
Komendantskaya E, McCusker G, Power J (2011) {Coalgebraic Semantics for
  Parallel Derivation Strategies in Logic Programming}. International
  Conference on Algebraic Methodology and Software Technology - AMAST
  13:111--127

\bibitem[{Komendantskaya et~al(2016)Komendantskaya, Power, and
  Schmidt}]{Komendantskaya16}
Komendantskaya E, Power J, Schmidt M (2016) {Coalgebraic Logic Programming:
  from Semantics to Implementation}. Journal of Logic and Computation 26(2):745
  -- 783

\bibitem[{Kowalski(1979)}]{Kowalski1979}
Kowalski R (1979) {Algorithm = Logic + Control}. Communications of the ACM
  22(7):424--436

\bibitem[{Kowalski(1986)}]{Kowalski1986}
Kowalski R (1986) {Logic for Problem-Solving}. North-Holland Publishing Co.

\bibitem[{Kowalski and Kuehner(1971)}]{KOWALSKI1971227}
Kowalski R, Kuehner D (1971) Linear resolution with selection function.
  Artificial Intelligence 2(3):227--260. \doi{10.1016/0004-3702(71)90012-9}

\bibitem[{Kripke(1965)}]{kripke1965semantical}
Kripke SA (1965) {Semantical Analysis of Intuitionistic Logic I}, vol~40,
  Elsevier, pp 92--130

\bibitem[{Lambek(1980)}]{lambek1980lambda}
Lambek J (1980) {From $\lambda$-calculus to Cartesian Closed Categories}. To H
  B Curry: Essays on Combinatory Logic, Lambda Calculus and Formalism pp
  375--402

\bibitem[{Lemmon(1978)}]{lemmon1978beginning}
Lemmon EJ (1978) {Beginning Logic}. Hackett Publishing

\bibitem[{Makinson(2014)}]{makinson2014inferential}
Makinson D (2014) {On an Inferential Semantics for Classical Logic}. Logic
  Journal of IGPL 22(1):147--154

\bibitem[{Martin-L{\"o}f(1975)}]{martin1975intuitionistic}
Martin-L{\"o}f P (1975) {An Intuitionistic Theory of Types: Predicative Part }.
  Studies in Logic and the Foundations of Mathematics 80:73--118

\bibitem[{Miller(1989)}]{miller1989logical}
Miller D (1989) {A Logical Analysis of Modules in Logic Programming}. The
  Journal of Logic Programming 6(1-2):79--108

\bibitem[{Miller et~al(1991)Miller, Nadathur, Pfenning, and
  Scedrov}]{miller1991uniform}
Miller D, Nadathur G, Pfenning F, et~al (1991) {Uniform Proofs as a Foundation
  for Logic Programming}. Annals of Pure and Applied logic 51(1-2):125--157

\bibitem[{Milner(1983)}]{milner1983calculi}
Milner R (1983) {Calculi for Synchrony and Asynchrony}. Theoretical Computer
  Science 25(3):267--310

\bibitem[{Milner(1984)}]{milner1984tactics}
Milner R (1984) {The Use of Machines to Assist in Rigorous Proof}.
  Philosophical Transactions of the Royal Society of London Series A,
  Mathematical and Physical Sciences 312(1522):411--422

\bibitem[{Milner(1989)}]{milner1989communication}
Milner R (1989) {Communication and Concurrency}. Prentice Hall

\bibitem[{Nascimento(2023)}]{nascimentothesis}
Nascimento V (2023) Foundational studies in proof-theoretic semantics. PhD
  thesis, Universidade do Estado do Rio de Janeiro

\bibitem[{Nascimento et~al(2023)Nascimento, Pereira, and
  Pimentel}]{nascimento2023ecumenical}
Nascimento V, Pereira LC, Pimentel E (2023) {An Ecumenical view of
  Proof-theoretic Semantics}. arXiv:230603656 Accessed July 2024

\bibitem[{Negri(2005)}]{Negri2005}
Negri S (2005) {Proof Analysis in Modal Logic}. Journal of Philosophical Logic
  34(5):507--544

\bibitem[{{nLab}(2021)}]{nLabFF}
{nLab} (2021) {free functor}.
  {\url{https://ncatlab.org/nlab/show/free+functor}. Access\-ed July 2024}

\bibitem[{O'Hearn and Pym(1999)}]{o1999logic}
O'Hearn PW, Pym DJ (1999) {The Logic of Bunched Implications}. Bulletin of
  Symbolic Logic 5(2):215--244

\bibitem[{Paulson(1994)}]{Paulson1994}
Paulson LC (1994) Isabelle: A Generic Theorem Prover. Springer,
  \doi{10.1007/BFb0030541}

\bibitem[{Pavlovi\'c(1990)}]{Pavlovic1990}
Pavlovi\'c D (1990) {Predicates and Fibrations}. PhD thesis, University of
  Utrecht

\bibitem[{Pfenning and Sch{\"u}rmann(1999)}]{Pfenning1999}
Pfenning F, Sch{\"u}rmann C (1999) System description: Twelf --- a meta-logical
  framework for deductive systems. In: Automated Deduction --- CADE, pp
  202--206

\bibitem[{Piecha(2016)}]{Piecha2016completeness}
Piecha T (2016) {Completeness in Proof-theoretic Semantics}. Advances in
  Proof-theoretic Semantics pp 231--251

\bibitem[{Piecha and Schroeder-Heister(2019)}]{Piecha2019incompleteness}
Piecha T, Schroeder-Heister P (2019) {Incompleteness of Intuitionistic
  Propositional Logic with Respect to Proof-theoretic Semantics}. Studia Logica
  107(1):233--246

\bibitem[{Piecha et~al(2015)Piecha, de~Campos~Sanz, and
  Schroeder-Heister}]{Piecha2015failure}
Piecha T, de~Campos~Sanz W, Schroeder-Heister P (2015) {Failure of Completeness
  in Proof-theoretic Semantics}. Journal of Philosophical Logic 44(3):321--335

\bibitem[{Plaisted and Zhu(1997)}]{Plaisted1997}
Plaisted DA, Zhu Y (1997) {The Efficiency of Theorem Proving Strategies}.
  Springer

\bibitem[{Plotkin(1981)}]{plotkin1981structural}
Plotkin GD (1981) {A Structural approach to Operational Semantics}. Tech. rep.,
  Aarhus University

\bibitem[{Portoraro(2024)}]{sep-reasoning-automated}
Portoraro F (2024) {Automated Reasoning}. In: Zalta EN, Nodelman U (eds) The
  {Stanford} Encyclopedia of Philosophy, {S}pring 2024 edn. Metaphysics
  Research Lab, Stanford University

\bibitem[{Pym and Ritter(2004)}]{pym2004reductive}
Pym DJ, Ritter E (2004) {Reductive Logic and Proof-search: Proof Theory,
  Semantics, and Control}, Oxford Logic Guides, vol~45. Oxford University Press

\bibitem[{Pym and Wallen(1993)}]{PW1993logic}
Pym DJ, Wallen LA (1993) Logic programming via proof-valued computations. In:
  Proceedings of the UK Conference on Logic Programming --- ALPUK. Springer, pp
  253--262

\bibitem[{Pym et~al(2024)Pym, Ritter, and Robinson}]{pym2024categorical}
Pym DJ, Ritter E, Robinson E (2024) Categorical proof-theoretic semantics.
  Studia Logica pp 1--38. \doi{https://doi.org/10.1007/s11225-024-10101-9}

\bibitem[{Reitzig(2012)}]{Reitzig}
Reitzig R (2012) {Discussion on Computer Science Stack Exchange.}
  \url{https://cs.stackexchange.com/questions/525/what-is-coinduction}.
  Accessed June 2024

\bibitem[{Rutten(2000)}]{RUTTEN20003}
Rutten J (2000) {Universal Coalgebra: a theory of systems}. Theoretical
  Computer Science 249(1):3--80. \doi{10.1016/S0304-3975(00)00056-6}

\bibitem[{Sandqvist(2005)}]{Sandqvist2005inferentialist}
Sandqvist T (2005) {An inferentialist Interpretation of Classical Logic}. PhD
  thesis, Uppsala University

\bibitem[{Sandqvist(2009)}]{Sandqvist2009CL}
Sandqvist T (2009) {Classical logic without Bivalence}. Analysis 69(2):211--218

\bibitem[{Sandqvist(2015)}]{Sandqvist2015IL}
Sandqvist T (2015) {Base-extension Semantics for Intuitionistic Sentential
  Logic}. Logic Journal of the IGPL 23(5):719--731

\bibitem[{{Saunders MacLane}({1971})}]{MacLane71}
{Saunders MacLane} ({1971}) {Categories for the Working Mathematician}.
  {Springer}

\bibitem[{Schroeder-Heister(1984)}]{Schroeder1984natural}
Schroeder-Heister P (1984) A natural extension of natural deduction. The
  Journal of Symbolic Logic 49(4):1284--1300

\bibitem[{Schroeder-Heister(2006)}]{Schroeder2006validity}
Schroeder-Heister P (2006) {Validity Concepts in Proof-theoretic Semantics}.
  Synthese 148(3):525--571

\bibitem[{Schroeder-Heister(2008)}]{Schroeder2007modelvsproof}
Schroeder-Heister P (2008) {Proof-Theoretic versus Model-Theoretic
  Consequence}. The Logica Yearbook 2007

\bibitem[{Schroeder-Heister(2018)}]{SEP-PtS}
Schroeder-Heister P (2018) {Proof-theoretic Semantics}. {The Stanford
  Encyclopedia of Philosophy}

\bibitem[{Seely(1983)}]{seely1983hyperdoctrines}
Seely RA (1983) {Hyperdoctrines, Natural Deduction and the Beck condition}.
  Mathematical Logic Quarterly 29(10):505--542

\bibitem[{Simon et~al(2007)Simon, Bansal, Mallya, and Gupta}]{simon2007co}
Simon L, Bansal A, Mallya A, et~al (2007) {Co-Logic Programming: extending
  logic programming with coinduction}. In: International Colloquium on
  Automata, Languages and Programming --- ICALP. Springer, pp 472--483

\bibitem[{Smullyan(1968)}]{Smullyan1968}
Smullyan RM (1968) Analytic Tableaux, Springer, pp 15--30.
  \doi{10.1007/978-3-642-86718-7_2}

\bibitem[{Stafford(2021)}]{Stafford2021}
Stafford W (2021) {Proof-theoretic Semantics and Inquisitive Logic}. Journal of
  Philosophical Logic

\bibitem[{Stafford and Nascimento(2023)}]{stafford2023}
Stafford W, Nascimento V (2023) {Following all the Rules: Intuitionistic
  Completeness for Generalized Proof-theoretic Validity}. Analysis
  \doi{10.1093/analys/anac100}

\bibitem[{Streicher(1988)}]{Streicher1988}
Streicher T (1988) {Correctness and Completeness of a Categorical Semantics of
  the Calculus of Constructions}. PhD thesis, University of Passau

\bibitem[{Szabo(1969)}]{Gentzen}
Szabo ME (ed)  (1969) {The Collected Papers of Gerhard Gentzen}. North-Holland
  Publishing Company

\bibitem[{Tarski(1936)}]{tarski1936concept}
Tarski A (1936) {On the Concept of Logical Consequence}. Logic, semantics,
  metamathematics 52:409--420

\bibitem[{Tarski(1955)}]{Tarski1955}
Tarski A (1955) {A Lattice-theoretical Fixpoint Theorem and Its Applications}.
  Pacific Journal of Mathematics 5(2):285--309

\bibitem[{Troelstra and Schwichtenberg(2000)}]{troelstra2000basic}
Troelstra AS, Schwichtenberg H (2000) {Basic Proof Theory}, Cambridge Tracts in
  Theoretical Computer Science, vol~43. Cambridge University Press

\bibitem[{Turi and Plotkin(1997)}]{turi97}
Turi D, Plotkin GD (1997) {Towards a Mathematical Operational Semantics}. In:
  Logic in Computer Science --- LICS. IEEE, pp 280--291,
  \doi{10.1109/LICS.1997.614955}

\bibitem[{Wallen(1987)}]{wallen}
Wallen LA (1987) Automated proof search in non-classical logics : efficient
  matrix proof methods for modal and intuitionistic logics. PhD thesis,
  University of Edinburgh

\bibitem[{Wansing(2000)}]{wansing2000idea}
Wansing H (2000) {The Idea of a Proof-theoretic Semantics and the Meaning of
  the Logical Operations}. Studia Logica 64:3--20

\bibitem[{Wikipedia(Accessed November 2024)}]{wikiproofassistants}
Wikipedia (Accessed November 2024) {Proof Assistant}.
  \url{https://en.wikipedia.org/wiki/Proof_assistant}

\bibitem[{Worrell(2005)}]{WORRELL2005184}
Worrell J (2005) {On the Final Sequence of a Finitary Set Functor}. Theoretical
  Computer Science 338(1):184--199.
  \doi{https://doi.org/10.1016/j.tcs.2004.12.009}

\end{thebibliography}

\end{document}